\RequirePackage{fix-cm}
\documentclass[twocolumn,epjc3]{svjour3}  
\smartqed  
\RequirePackage{graphicx}
\RequirePackage{bm}
\RequirePackage{amssymb}
\RequirePackage[final]{changes}

\newcommand*\mean[1]{\langle #1 \rangle}

\journalname{Eur. Phys. J. E}
\begin{document}

\title{Strain localization in planar shear of granular media: The role of porosity and boundary conditions
}


\author{Stanislav Parez\thanksref{e1,addr1,addr2}
        \and
        Tereza Travnickova\thanksref{addr1}
        \and
        Martin Svoboda\thanksref{addr1}
        \and
        Einat Aharonov\thanksref{addr3}                
}

\thankstext{e1}{e-mail: parez@icpf.cas.cz}


\institute{Czech Academy of Sciences, Institute of Chemical Process Fundamentals, Prague, Czech Republic \label{addr1}
           \and
           Jan Evangelista Purkyn\v{e} University in Ústí nad Labem, Faculty of Science, Ústí nad Labem, Czech Republic \label{addr2}
           \and
           Institute of Earth Sciences, Hebrew University of Jerusalem, Israel \label{addr3}
}

\date{Received: date / Accepted: date}

\abstractdc{
Shear strain localization into shear bands is associated with velocity weakening instabilities and earthquakes. Here, we simulate steady-state plane-shear flow of numerical granular material (gouge), confined between \deleted{rough, }parallel surfaces. Both constant shear stress and constant strain-rate boundary conditions are tested and the two types of boundary conditions are found to yield distinct velocity profiles and friction laws. The inertial number, $I$, exerts the largest control on the layers' behavior, but additional dependencies of friction on \replaced{normal stress}{grain stiffness} and thickness of the layer are observed under constant stress boundary condition. We find that shear-band localization, which is present in the quasistatic regime ($I<10^{-3}$) in rate-controlled shear, is absent under stress-controlled loading. In the latter case, flow ceases when macroscopic friction coefficient approaches the quasistatic friction value. The inertial regime that occurs at higher inertial numbers ($I>10^{-3}$) is associated with distributed shear, and friction and porosity that increase with shear rate (rate-strengthening regime). The finding that shear under constant stress boundary condition produces the inertial, distributed shear but never quasistatic, localized deformation is rationalized based on low fluctuations of shear forces in granular contacts for stress-controlled loading. By examining porosity within and outside a shear band, we also provide a mechanical reason why the transition between quasistatic and inertial shear coincides with the transition between localized and distributed strain.
}

\maketitle

%
%
\section{Introduction}
\label{intro}
Shear of granular media controls many industrial and geological settings. For rocks, granular shear occurs within geological fault zones and in landslides. During shear and sliding, faults experience wear and accumulate an increasingly thick layer of crushed grains (termed fault gouge), which becomes the locus of sliding, \textit{e.g.,} \cite{Scholz1987,Chester1993,Billi2005}. Field observation of fault zones \cite{Arboleya1995,Cashman2000,Hayman2004,Boullier2009,Shalev2013,Smeraglia2017}, laboratory experiments of shearing confined grains \cite{Logan1979,Marone1990,LOGAN1992,Beeler1996,Spiers2007,Reches2010,DiToro2014,Mitchell2016} and numerical simulations \cite{Mora1999,Aharonov2002,Morgan1999a,MAIR2008} show that shear strain often localizes into discrete, planar zones that constitute gouge layers. Also in landslides, when the slide is thick enough, most of the shear is concentrated at its base, in a localized shear band \cite{Li2021}.

Results from laboratory experiments on analog fault zones suggest that strain localization within granular layers also coincides with the transition to velocity weakening and unstable sliding, manifested by stick-slip behavior. This transition for geological fault zones is in turn identified with the transition from stable creep to unstable earthquakes \cite{Logan1979,Marone1990,LOGAN1992,Beeler1996}, marking the question of localization as a fundamental aspect of earthquake physics \cite{Ben-Zion2003,Marone1991}. The transition between stable and unstable sliding has been found to be predominantly controlled by mineral composition and structure \cite{Brace1972,Ikari2011,Spiers2007}, effective confining stress \cite{Brace1972,Shimamoto1986}, stress path \cite{French2016,Wu2013}, pore fluid effects \cite{Marone2009,Faulkner2018,French2016,GOREN2009}, temperature effects \cite{Chester1994,Mitchell2016,Aharonov2018} and time-dependent chemical processes at grain contacts \cite{Frye2002}. Here we leave thermo-, hydro- and chemical effects aside and seek purely mechanical origin for this transition and conditions under which constitutive laws of granular media allow localization.

The physics of localization onset in continuum mechanics has been the subject of long research. A general theory, \textit{e.g.,} \cite{Rice1975,Rice1976,VARDOULAKIS1976,Vardoulakis1980,Muhlhaus1987,Sulem1990,LARSSON1996,Weir2003,Einav2006}, views localization as a bifurcation point for which constitutive equations of the rate-boundary-value problem change their type from elliptic (prior to localization) to hyperbolic \cite{Vardoulakis1995}. Conditions for the onset of shear band localization can be predicted if the constitutive equations of the material (prior to localization) feature bifurcation for which a solution in the form of localized deformation exists. Rudnicki and Rice \cite{Rice1975} predicted both the shear band orientation and the amount of strain accommodated within the shear band for realistic constitutive relations for brittle rocks under compressive principal stresses. Later studies have analyzed effects of pore fluid pressure, shear heating and chemical reactions \cite{rice2006,Sulem2009,Veveakis2011,Sulem2012,Sulem2013,Rice2014}.

Experimental work in soil mechanics \cite{Desrues1996} and the associated theory of the critical state \cite{Lambe1969} point at a close connection between localization and porosity. The experiments of Desrues \textit{et al.} \cite{Desrues1996} (see their Fig.~19) demonstrate that porosity may be tweaked to control localization: shearing over-consolidated sand localizes shear into shear bands, while under-consolidated samples compact diffusely, both cases shearing finally at a porosity value which is independent of the initial condition, the critical porosity. The critical value is attained only within the shear band in the over-consolidated sample. 

In dry granular media, such as sand, steady-state porosity and also friction are observed to be functions of normal stress and shear rate via the dimensionless inertial number \cite{GDRMiDi,Forterre2008,daCruz2005,Singh2015}, which is proportional to the shear rate and inversely proportional to the square root of the normal stress. The inertial number measures the relative importance of grain inertia to stress forces; larger inertial numbers produce increased agitation of grains. As the inertial number increases, \textit{e.g.,} by increasing shear rate, both porosity and friction coefficient increase. In this so-called inertial regime, friction becomes clearly rate strengthening as a result of increased dissipation in granular collisions and work against confining stress invoked by the dilatant behavior. The strengthening stabilizes flow, producing uniformly distributed shear rate in planar shear configuration, and facilitates approach to steady shear in transient flows \cite{Parez2015,Parez2016,Shojaaee2012a}.

Localization has been mapped to emerge under low slip rates and high normal stresses \cite{Aharonov2002,Shojaaee2012a,Li2021}, corresponding to low inertial numbers (quasistatic regime). In this quasistatic regime, friction and porosity attain seemingly constant (independent of the inertial number or shear rate), low values. However, some authors argue that this regime is in fact slightly rate-weakening \cite{DeGiuli2017,Dijksman2011,Kuwano2013}. Such weakening may rationalize initiation of shear-band localization, as well as hysteresis of the angle of repose observed in granular avalanches down inclined planes \cite{Pouliquen_book}. DeGiuli and Wyart \cite{DeGiuli2017} proposed that the origin of the weakening is an acoustic noise induced by collisions between grains. The noise can trigger slip on contacts sufficiently close to the sliding threshold, in a similar way originally proposed within the model of acoustic fluidization \cite{Melosh1979}. Activation of these contacts amplifies slip and produces more acoustic noise, promoting an instability. Barker and Gray \cite{barker_gray_2017} investigated stability of the inertial-number controlled rheology, similarly to the approach adopted by Rudnicki and Rice \cite{Rice1975} for brittle plasticity. They found that at sufficiently low inertial numbers the rheology becomes ill-posed, amplifying small-wavelength perturbations. This instability may either be viewed as the onset of shear-band localization or attributed to the empirical nature of the functional form describing friction coefficient as a function of the inertial number. Interestingly, the instability forms even in the incompressible flow approximation and therefore is not driven by changes in porosity. Apparently, there is much controversy regarding the question of the relative roles of porosity and granular rheology in controlling strain localization, mostly due to the fact that mechanisms driving localization in dry granular media are not well understood and lack solid quantitative description.

In this paper, we study, using numerical discrete element simulations, the effect of boundary conditions on the distribution of strain and porosity in granular layers subject to planar shear. We compare constant shear rate and constant shear stress boundary conditions and test the ability of the two boundary conditions to localize strain into shear bands. In agreement with the vast majority of experiments and numerical simulations, shear driven by controlled shear rate indeed allows for strain localization into planar structures \cite{LOGAN1992,Beeler1996,DiToro2014,MAIR2008,daCruz2005,Aharonov2002}. On the other hand, studies of shear under stress control are extremely rare. An exception is the work of Clark \textit{et al.} \cite{Clark2018}. They measured the amount of shear strain before flow cessation as a function of applied shear stress and system size, but do not report on strain localization. In this work, we find that shear-band localization, which is present in the quasistatic regime in the rate-controlled shear, is absent for the stress-controlled loading. In the case of constant applied stress, flow ceases when macroscopic friction coefficient approaches the quasistatic friction value. We rationalize this finding based on low fluctuations of shear forces in granular contacts, leading to a reduced mechanical noise.

An additional major finding of this paper is related to a mechanism controlling strain localization based on critical porosity. The mechanism is derived from energy balance following Frank and later studies \cite{Frank1965,Marone1990}. The theory relates strength of a confined layer to its dilatancy: dilatancy provides strengthening of the layer as a result of work expended in effecting a volume change against the confining stress. While dilatancy is necessary for the onset of shear in compacted layers \cite{Reynolds1885}, the energetic penalty is minimized when the system dilates only locally, within a shear band, to reach the critical porosity, whereas the spectator regions have sub-critical porosity. This is consistent with the triaxial compression tests on sand \cite{Desrues1996} (see their Fig.~19), where over-consolidated samples reach critical porosity only within a shear band, while the global porosity is lower. Since the global porosity is an increasing function of the inertial number, reflecting the level of mechanical noise in the system, the critical porosity can be reached globally throughout the layer for a large enough inertial number. Such conditions mark the transition to the distributed shear regime, when localization fully vanishes.

\section{Numerical procedures\deleted{ and units}}
\label{sec:numerics}
\begin{figure}
\includegraphics[width=0.5\textwidth]{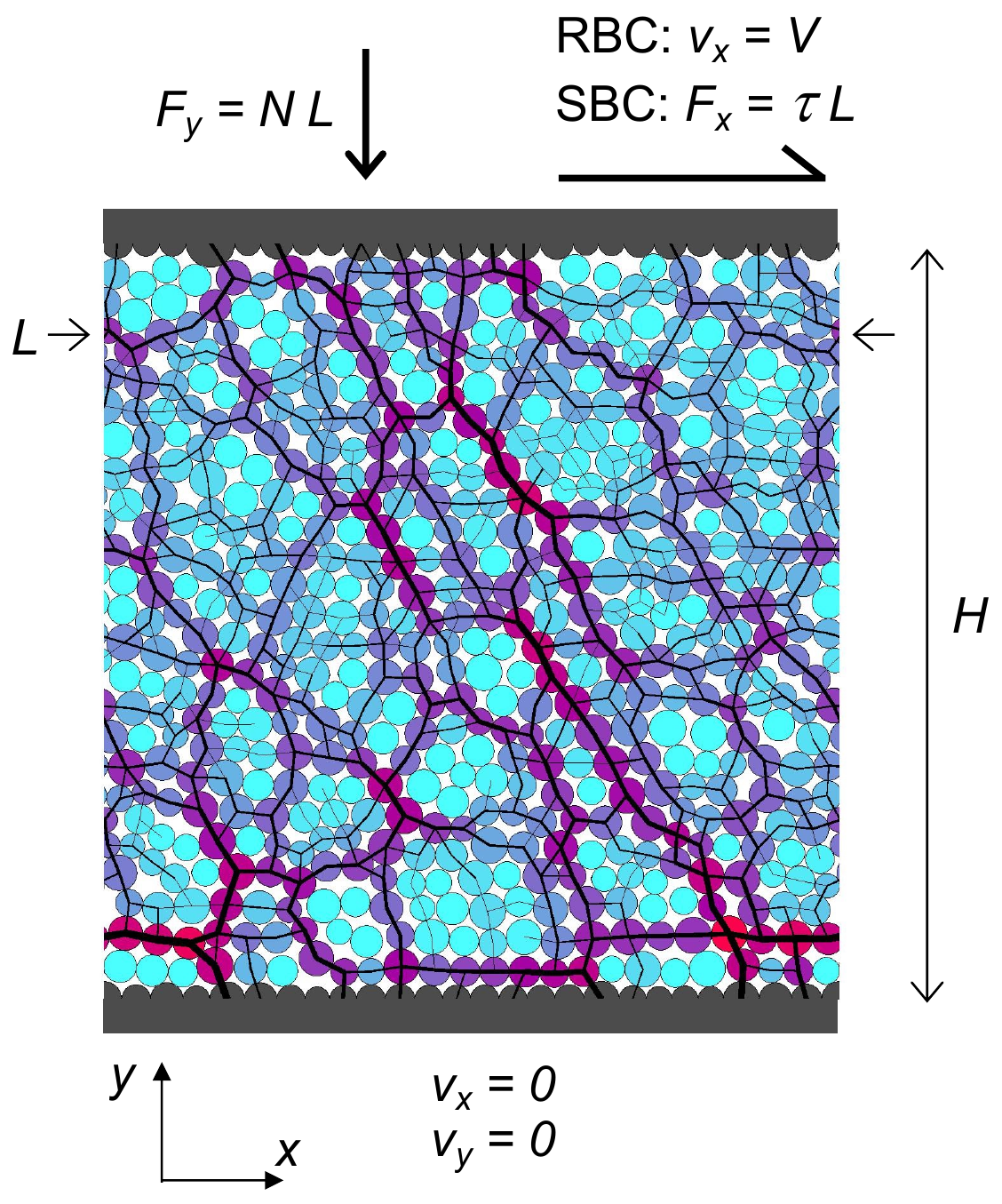}	
\caption{Numerical representation of granular shear. A dense assembly of dissipative frictional disks is sandwiched between two \replaced{parallel}{rough} walls made of glued disks. The bottom wall is static, while the top wall is subject to normal stress $N$. Shear motion is driven either by constant $x$-velocity $V$ of the top wall (RBC), or by constant shear stress $\tau$ applied on the top wall (SBC). The system is periodic in $x$ direction. Colors and line widths indicate magnitude of the normal force between grains (red grains connected by thick lines are most stressed).}
\label{fig:schematic}       
\end{figure}

We employ discrete element method (DEM) \cite{DEM} to simulate planar shear of a granular layer under constant applied shear stress or shear strain rate. The layer is confined in the vertical direction by two parallel\replaced{, granular}{ rough} surfaces, as in Fig.~\ref{fig:schematic}. 

Grains are modeled as two-dimensional (2D) disks with rotational and translational degrees of freedom. Disks interact via pair-wise contact forces according to a linear spring model with velocity-dependent damping and Coulomb friction criterion. Let $R_i$, $m_i$, $\mathbf v_i$, $\bm \omega_i$ and $\mathbf r_i$ be radius, mass, linear and angular velocity, and position vector of the center of grain $i$, respectively. Grains $i$ and $j$ interact only when they form a contact, defined by the condition that the two disks overlap, $R_i+R_j-r_{ij}>0$, where $r_{ij}$ is the length of the vector $\mathbf r_{ij}=\mathbf r_{j}-\mathbf r_{i}$ connecting disks' centers. Normal and tangential forces on grain $i$ due to interaction with grain $j$, $\mathbf F_{n_{ij}}$ and $\mathbf F_{t_{ij}}$, respectively, are given by
\begin{eqnarray}
\label{normal_force}
\mathbf F_{n_{ij}} &=& -k_n \delta_{ij} \mathbf{n}_{ij}  + \gamma_n m_\mathrm{eff} \mathbf{v}_{n_{ij}}  \,,  \\
\label{tangential_force}
\mathbf F_{t_{ij}} &=& -k_t \lambda_{ij} \mathbf{t}_{ij}  + \gamma_t m_\mathrm{eff} \mathbf{v}_{t_{ij}} \,, \qquad \mathrm{if} \quad F_{t_{ij}} < \mu_{g}  F_{n_{ij}} \,,
\end{eqnarray}
in which $k_{n,t}$ and $\gamma_{n,t}$ are elastic and viscoelastic constants (material parameters), respectively, $\mathbf n_{ij}  =\mathbf r_{ij}/r_{ij}$ is the unit vector in the normal direction to the contact, $\mathbf{t}_{ij}$ is the unit vector in the tangential direction, $\mathbf{v}_{n_{ij}}=(\mathbf v_j - \mathbf v_i)\cdot \mathbf n_{ij}\, \mathbf n_{ij}$ is relative normal velocity, $\mathbf v_{t_{ij}}=\mathbf v_j - \mathbf v_i - \mathbf{v}_{n_{ij}} + (R_i \bm \omega_i + R_j \bm \omega_j)\times \mathbf n_{ij}$ is relative tangential velocity, $m_\mathrm{eff}=m_i m_j/(m_i + m_j)$ is the effective mass, $\delta_{ij}=R_i+R_j-r_{ij}$ is the contact deflection (overlap between disks), and $\lambda_{ij}=\int v_{t_{ij}} \mathrm d t$ is the tangential displacement of the two contacting grains, where the integration over time $t$ runs from the time of formation of the contact. 

The magnitude of the tangential force in Eq.~(\ref{tangential_force}) is truncated once the Coulomb friction criterion is met, $F_{t_{ij}} = \mu_{g}  F_{n_{ij}}$, where $\mu_{g}$ is grain surface friction coefficient. At that moment, the tangential force is switched to the frictional force $F_{t_{ij}} = \mu_{g}  F_{n_{ij}}$, independent of elastic or viscoelastic contributions. The truncation of the resisting tangential force allows the grains to slip past that contact.

\deleted{Our results are given in non-dimensional units using length, mass and time scales of mean grain diameter, $d$, mass of the mean-size grain, $m$, and time scale of propagation of an elastic wave through the mean-size grain, $t_0=\sqrt{m / k_n}$, respectively. All other quantities are scaled by proper functions of $d$, $m$ and $t_0$ to match their dimensions.
 
The following DEM parameters were used, similarly to previous studies of 2D systems \cite{daCruz2005,Aharonov2002}: $k_n=1$, $k_t=0.5$, $\gamma_n=0.5$, $\gamma_t=0$, $ \mu_{g}=0.5$. The distribution of grain sizes is derived from Gaussian distribution with $\sigma=1$ while keeping the maximum polydispersity of $\pm 20\%$. Polydispersity prevents crystallization of the system. The coefficient of restitution, which characterizes inelasticity of collisions, is $\approx 0.3$ for the selected parameters. Nevertheless, the dominant mechanism for energy dissipation in dense granular flows is work of the friction force $F_{t_{ij}} = \mu_{g}  F_{n_{ij}}$ \cite{daCruz2005}. On that basis, we neglect the viscoelastic damping in the tangential force by setting $\gamma_t=0$.}

Forces defined by Eqs.~(\ref{normal_force}) -- (\ref{tangential_force}) are plugged into Newtonian equations of motion to solve linear and angular momenta for each grain; more details can be found in Refs. \cite{DEM,Frenkel2002,daCruz2005}. The velocity Verlet integrator \cite{Frenkel2002} was used to propagate the equations of motions with a time step of \replaced{$0.1$ of the time required for a sound wave to travel across the smallest grain in the system. This time step is small enough to resolve elastic waves due to particle collisions, which represent the fastest energy and momentum transfer in the system.}{$0.1$, small enough to resolve elastic waves due to particle collisions.}

Grains are packed into a layer confined in the $y$ direction by two parallel\deleted{, rough} surfaces. The surfaces are constructed from grains glued together into a linear array. The size distribution of the surface grains is the same as for the interior grains, and interactions between the surface and the interior grains are governed by the same contact forces, Eqs.~(\ref{normal_force}) -- (\ref{tangential_force}), using the same mechanical parameters. \added{Our choice of boundaries sits between perfectly smooth, planar walls, which promote boundary slip, and very rough walls, \textit{e.g.,} constructed of grains separated by gaps between them \cite{Shojaaee2012b}, which impose no-slip boundary conditions.} Periodic boundary conditions are applied in the $x$ (flow) direction to model spatially extensive and homogeneous deformation along the $x$ axis. \deleted{Thickness of the layer, $H$, which directly affects shear rate across the layer, was varied between $H=24-200$. The domain length in periodic direction, $L$, was maintained at $L=96$; variation of $L$ between $48-120$ resulted in no significant effect on the studied structural and rheological characteristics.}

The bottom surface is fixed, while the top surface is pushed against the layer by normal stress, $N$. The normal stress is applied by exerting an external force $F_y= - N L$ onto the top surface. Shear flow of the layer is driven by two distinct types of boundary conditions. The first type is a constant rate boundary condition (RBC), in which the top surface moves with constant velocity $V$ in $x$ direction. The second type of the boundary condition is a constant stress boundary condition (SBC). Here, a constant shear stress $\tau$ is applied by applying an external force $F_x=\tau L$ onto the top surface. Note that instantaneous normal and shear stress fluctuates even under SBC due to the fluctuating forces exerted by grains in contact with the boundary. Gravity and pore fluid effects are absent in this work.

The friction coefficient was calculated by averaging local stress tensor, $\sigma(\mathbf{r})$, over the entire system, $\mu=\mean{\sigma_{xy}(\mathbf{r})/\sigma_{yy}(\mathbf{r})}$. The stress tensor is calculated from grain configurations as described in \textit{e.g.,} \cite{daCruz2005,Singh2015}. Nevertheless, almost identical values were obtained by calculating the friction coefficient as the ratio of shear and normal stresses measured at the boundaries, because the stress tensor $\sigma(\mathbf{r})$ is homogeneous in the steady planar shear flow\added{, apart from deviations that arise for smooth boundaries \cite{Shojaaee2012b}}.

\deleted{The applied normal stress, shear velocity and shear stress ranged $N=10^{-7}-10^{-3}$, $V=10^{-5}-10^{-3}$ and $\tau=(0.25 - 0.4) N$, respectively. To convert these values into real units, material parameters need to be plugged in. For quartz grains with Young's modulus $6\cdot10^{10}$ Pa, density $2640$ kg\,m$^{-3}$ and grain size of $10^{-3}$ m (resulting in normal stiffness $\hat{k}_n \approx 6\cdot10^{7}$ N\,m$^{-1}$), the corresponding ranges of normal stress and slip rate are $\hat{N}=6\cdot(10^{3}-10^{7})$ Pa and $\hat{V}=6.6\cdot (10^{-2}-1)$ m\,s$^{-1}$. The range of applied normal stress reflects realistic values typical for landslides and not too deep fault zones, \textit{e.g.} \cite{Smeraglia2017,Marone1990,LOGAN1992}.}

Samples are initialized as dense packings with solid fraction close to the random close packing. This is achieved by pre-shearing at a low velocity $V=10^{-6}$ followed by a static relaxation. Simulations are allowed sufficient time for a steady flow to develop. To obtain steady-flow time averages, measurements from the transient period are discarded. Local and single-grain quantities are averaged over surface-parallel layers \replaced{2 grains thick}{of thickness 2}. The resulting averages are functions of $y$ only, reflecting homogeneity in the $x$ direction.

\section{Units, parameter ranges and dimensional analysis}
\label{sec:units}
\added{Our results are given in non-dimensional units using length, mass and force scales of mean grain diameter, $\hat{d}$, mass of the mean-size grain, $\hat{m}$, and grain normal stiffness coefficient times mean grain diameter, $\hat{k}_n \hat{d}$, respectively. All other quantities are scaled by proper functions of $\hat{d}$, $\hat{m}$ and \replaced{$\hat{k}_n$}{$t_0$} to match their dimensions, \textit{e.g.,} time is measured in units of $\hat{t}_0=\sqrt{\hat{m} / \hat{k}_n}$. Hat over a symbol indicates dimensional quantities, \textit{e.g.,} $\hat{H}$ is in [m], while $H=\hat{H}/\hat{d}$ is dimensionless.}

\begin{table*}
\caption{List of parameters and their values in non-dimensional units using length, mass and force scales of $\hat{d}$, $\hat{m}$ and $\hat{k}_n \hat{d}$, respectively. Hatted parameters ($\hat{\,}$) denote dimensional quantities.}
\label{tab:setup}       
\begin{tabular}{lll}
\hline\noalign{\smallskip}
Parameter & Symbol & Value or range  \\
\noalign{\smallskip}\hline\noalign{\smallskip}
tangential stiffness coefficient & $k_t=\hat{k}_t/\hat{k}_n$ & 0.5 \\
normal damping coefficient & $\gamma_n=\hat{\gamma}_n \sqrt{\hat{m}/\hat{k}_n}$ & 0.5 \\
tangential damping coefficient & $\gamma_t=\hat{\gamma}_t \sqrt{\hat{m}/\hat{k}_n}$ & 0 \\
surface friction coefficient & $\mu_g$ & 0.5 \\
polydispersity & & $\pm 20\%$ \\
layer thickness & $H=\hat{H}/\hat{d}$ & $24-200$ \\
normal stress or inverse contact stiffness number$^{a}$ & $N=1/\kappa=\hat{N}{\hat{d}}^{D-2}/\hat{k}_n$ & $10^{-7}-10^{-3}$ \\
apparent friction coefficient$^{b}$ & $\mu=\tau/N$ & $0.25 - 0.4$ \\
shear velocity$^{c}$ & $V=\hat{V}\sqrt{\hat{m} / \hat{k}_n\hat{d}^2}$ & $10^{-5}-10^{-3}$ \\
\noalign{\smallskip}\hline
\multicolumn{2}{l}{$^{a}$ See text below Eq.~(\ref{kappa}). $D$ is the spatial dimension.}\\
\multicolumn{2}{l}{$^{b}$ $\mu$ is controlled under SBC.}\\
\multicolumn{2}{l}{$^{c}$ $V$ is controlled under RBC.}
\end{tabular}
\end{table*}

\added{A list of material parameters and boundary conditions characterizing numerical setups is given in Table \ref{tab:setup}. Material parameters are chosen in close similarity to previous DEM studies of 2D systems \cite{daCruz2005,Aharonov2002}: $k_n=1$, $k_t=0.5$, $\gamma_n=0.5$, $\gamma_t=0$, $ \mu_{g}=0.5$. The coefficient of restitution, which characterizes inelasticity of collisions, is $\approx 0.3$ for the selected parameters. Nevertheless, the dominant mechanism for energy dissipation in dense granular flows is work of the friction force $F_{t_{ij}} = \mu_{g}  F_{n_{ij}}$ \cite{daCruz2005}. On that basis, we neglect the viscous damping in the tangential force by setting $\gamma_t=0$.

The distribution of grain sizes is derived from Gaussian distribution with $\sigma=1$ while keeping the maximum polydispersity of $\pm 20\%$. Polydispersity prevents crystallization of the system. Thickness of the layer, $H$, which directly affects shear rate across the layer, was varied between $H=24-200$. The domain length in periodic direction, $L$, was maintained at $L=96$; variation of $L$ between $48$ and $120$ resulted in no significant effect on the studied structural and rheological characteristics.

The applied normal stress, shear stress and shear velocity ranged $N=10^{-7}-10^{-3}$, $\tau=(0.25 - 0.4) N$ and $V=10^{-5}-10^{-3}$. In real units, illustrative ranges of normal stress and shear velocity are $\hat{N}=6\cdot(10^{3}-10^{7})$ Pa and $\hat{V}=6.6\cdot (10^{-2}-1)$ m\,s$^{-1}$, using grain size $\hat{d}=10^{-3}$ m, Young's modulus $\hat{E}=\hat{k}_n/\hat{d} =6\cdot10^{10}$ Pa and density $\hat{\rho}=6\hat{m}/\pi \hat{d}^3 = 2640$ kg\,m$^{-3}$ representative of quartz grains. The range of applied normal stress reflects realistic values typical for the geological setting of landslides and not too deep fault zones, \textit{e.g.,} \cite{Smeraglia2017,Marone1990,LOGAN1992}. Shear velocity corresponds to fault rupture nucleation and propagation slip rates.

The parameters listed in Table \ref{tab:setup} have been subject to numerous tests in previous studies to determine their effect on constitutive behavior. The effects of material parameters $k_t/k_n$, $\gamma_n$, $\gamma_t$ and $ \mu_{g}$ were found rather small for dense granular flows \cite{num_Silbert,daCruz2005}, except for the case of frictionless grains ($\mu_{g}=0$). The key parameter groups controlling the constitutive behavior were identified to be the inertial number and the contact stiffness number \cite{GDRMiDi,daCruz2005,Forterre2008,deCoulomb2017,Singh2015}. The inertial number, $I$, describes the ratio of inertial to stress forces, or, alternatively, the ratio of the inertial timescale ($\hat{d}\sqrt{\hat{\rho}/\hat{N}}$) and the macroscopic deformation timescale ($\hat{H}/\bar{\hat{v}}_x(\hat{H})$) 
\begin{equation}
\label{I}
I=\frac{\bar{\hat{v}}_x(\hat{H}) \, \hat{d}}{\hat{H}\sqrt{\hat{N}/\hat{\rho}}} \,,
\end{equation}  
where $\bar{\hat{v}}_x(\hat{H})$ is the average slip rate of the top wall and $\hat{\rho}$ is mass density of grains (in [kg\,m$^{-D}$], where $D$ is the spatial dimension). In the case of RBC, $\bar{\hat{v}}_x(\hat{H})$ is identical to the applied slip rate $\hat{V}$. Note that all quantities in Eq.~(\ref{I}) could also be non-dimensional (without the hats) because the product is dimensionless.

The contact stiffness number, $\kappa$, is proportional to the ratio of squares of the inertial timescale and the collision timescale ($\sqrt{\hat{m}/\hat{k}_n}$). For our choice of a stress scale, $\hat{k}_n / \hat{d}^{D-2}$ ($\approx$ Young's modulus $\hat{E}$ in 3D), the non-dimensional normal stress $N$ is identical with the inverse contact stiffness number
\begin{equation}
\label{kappa}
N=\frac{1}{\kappa} = \frac{\hat{N}\hat{d}^{D-2}}{\hat{k}_n}  \,.
\end{equation}  
Because of the linear elasticity model employed in this work, $N$ can be interpreted as the average compressive strain of a grain, $N \approx \hat{\delta}/\hat{d}$. This can be seen by balancing the average (compressive) normal-stress-induced force on a grain, $\hat{N} \hat{d}^{D-1}$, and the repulsive elastic force, $\hat{k}_n \hat{\delta}$, in Eq.~(\ref{normal_force}). Hence, $N$ represents contact softness, i.e. inverse contact stiffness: larger $N$ induces larger $\hat{\delta}/\hat{d}$, making grains effectively softer. 

When $N$ is varied such that $I$ is kept constant (\textit{e.g.,} by simultaneous change in $\hat{V}$ and $\hat{N}$ so that $\hat{V} \sim \sqrt{\hat{N}}$), constitutive behavior may still display a dependence on $N$ due to the variation in $\kappa$. We will conventionally refer to such changes in constitutive behavior as the effect of the contact stiffness number rather than the effect of normal stress. This is to prevent confusion with a simultaneous effect of the normal stress on the inertial number, according to Eq.~(\ref{I}), which often dominates the system's rheology. Note, however, that the normal stiffness coefficient of grains, $\hat{k}_n$, was kept constant among different setups in this work, while it was the normal stress, $\hat{N}$, that was actually varied, giving rise to changes in the ``contact stiffness." A similar situation is met in the field: stiffness of various geological grains varies only moderately, \textit{e.g.,} by a factor of 7 from halite to garnet, but closer to a factor of 2 for more common silicates, while the normal stress differs by several orders of magnitude among various tectonic loadings and burial depths of faults.
} 

\section{Constant rate boundary conditions}
\label{sec:RBC}

\begin{figure*}
\includegraphics{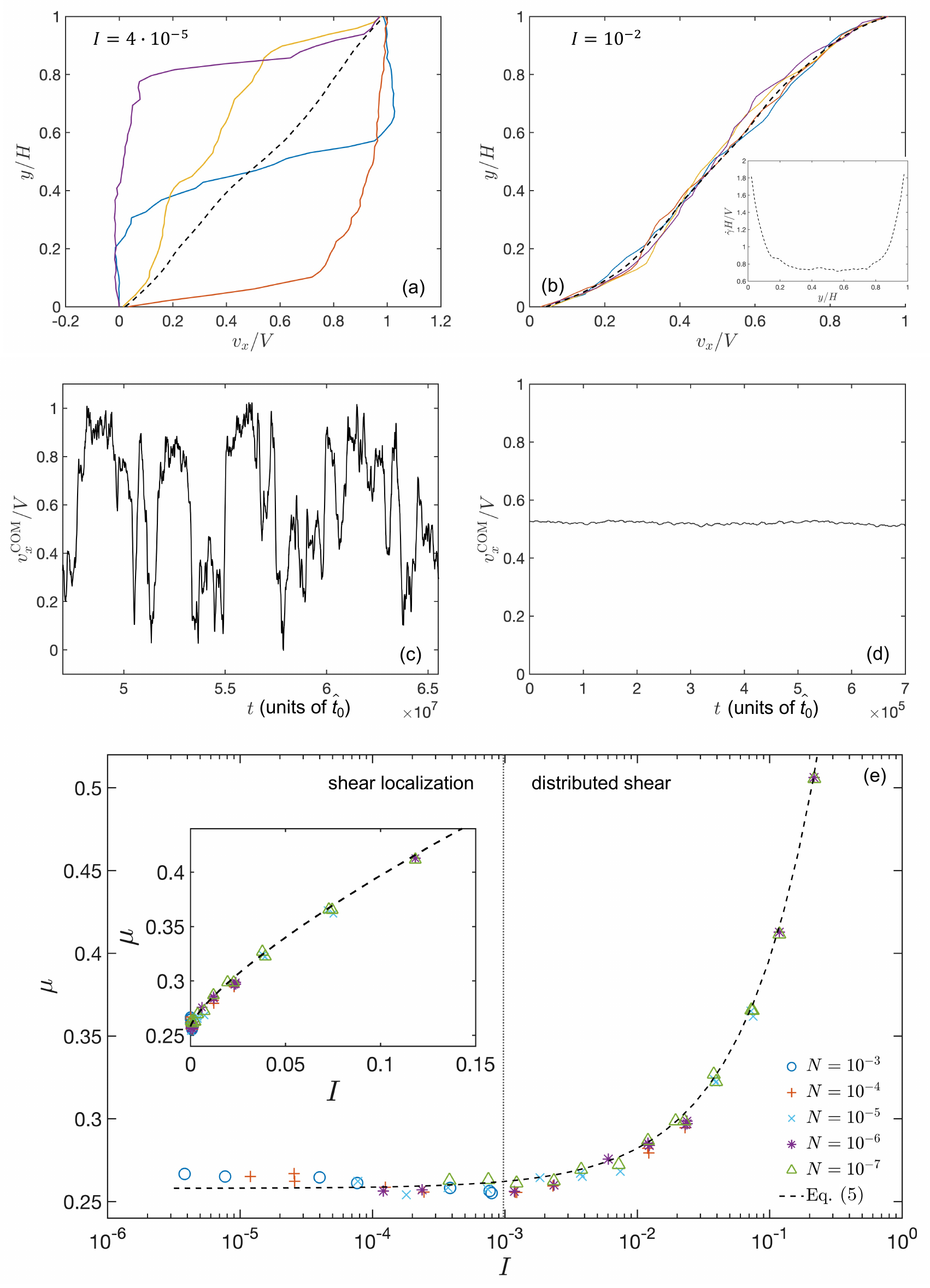}
\caption{Response of the layer under RBC. (a--b) Instantaneous shear-velocity profiles at random time instants show shear bands under $I<10^{-3}$ and distributed shear under $I>10^{-3}$. Thickness $H=96$ in both cases. The dashed line denotes steady-state time average. Inset of (b) displays the profile of shear strain rate, showing a deviation from a uniform value near boundaries. (c--d) Center-of-mass velocity as a function of time for the same conditions as in (a--b). (e) Friction coefficient \textit{vs.} inertial number. Different symbols represent different values of normal stress as given in the legend. The dashed line is Eq.~(\ref{mu-I}). Inset shows the same data in a linear scale.}
\label{fig:RBC}       
\end{figure*}

Shear driven by RBC is the most frequent choice of boundary conditions applied in experiments and simulations. The planar shear configuration studied here has been extensively addressed in the literature, \textit{e.g.,} \cite{daCruz2005,GDRMiDi,Forterre2008,Shojaaee2012a,DeGiuli2017}. In this section we review results of numerical simulations for this type of boundary conditions in order to make a comparison to SBC-driven shear in the next section.

Figure \ref{fig:RBC} shows local flow velocity, $v_x(y)$, evolution of the center-of-mass velocity, $v_x^\mathrm{COM}(t)$, and macroscopic friction coefficient, $\mu$, of a granular layer sheared under RBC. The response of the layer can be categorized into two regimes depending on a dimensionless parameter called the inertial number \deleted{
where $\bar{v}_x(H)$ is the average slip rate of the top wall and $\rho$ is mass density of grains. In the case of RBC, $\bar{v}_x(H)$ is identical to the applied slip rate $V$. The inertial number describes the ratio of inertial to stress forces, or, alternatively, the ratio of the inertial time scale ($d\sqrt{\rho/N}$) and the macroscopic deformation time scale ($1/\dot{\gamma}$) \cite{GDRMiDi,Forterre2008,daCruz2005}}.
The left panel of Fig.~\ref{fig:RBC} (a,c) displays results for relatively low $I=4 \cdot 10^{-5}$ using a setup with $V=10^{-4}$, $N=10^{-3}$ and $H=96$, while the right panel (b,d) displays results for high $I=10^{-2}$ using a setup with $V=10^{-3}$, $N=10^{-6}$ and $H=96$.

Under low inertial numbers, $I<10^{-3}$, imposed shear is accommodated in narrow shear bands (Fig.~\ref{fig:RBC}a). Each line represents instantaneous flow velocity profile at a random time step. Velocity varies significantly over a narrow interval of $y$ coordinate (shear band), about 20 grains thick. Position of the shear band changes randomly throughout the layer as documented by the linear time-averaged profile (dashed line). Change of the location of the shear band is accompanied by a transient diffuse shear (yellow line). Migrating shear bands are characteristic for shear with rough boundaries; smooth boundaries help localize shear at the boundaries \cite{Shojaaee2012a}. Thickness $H$ used to normalize the y-coordinate in the plots displaying velocity profiles is the steady-state thickness.

The center-of-mass shear velocity, $v_x^\mathrm{COM}$, varies intermittently with time (Fig.~\ref{fig:RBC}c) as a result of variation in the position of the shear band. Since velocity below the shear band's position is small and velocity above the shear band is close to $V$, different portions of the layer are mobilized as the shear band's position changes, leading to intermittent jumps in the evolution of $v_x^\mathrm{COM}$. The range of $v_x^\mathrm{COM}$ varies between $\approx 0$ and $\approx V$ corresponding to the shear band occurring at the moving and the fixed wall, respectively.

Finally, friction coefficient associated with shear localized within shear bands is $\mu \approx 0.26$ with little dependence on the applied shear velocity and normal stress (Fig.~\ref{fig:RBC}e). The value of friction coefficient in this regime is identified with yield strength, $\mu_\mathrm{qs}$. A subtle decreasing variation of friction with the inertial number for $I<10^{-3}$ was reported in Refs. \cite{DeGiuli2017,Kuwano2013}. This trend is somewhat apparent also in the data in Fig.~\ref{fig:RBC}e, albeit the prominence of the suggested minimum of friction at $I=10^{-3}$ is of a similar size as the uncertainty of the data, which is approximately equal to the size of the data points.

For high inertial numbers, $I>10^{-3}$, \textit{i.e.,} fast displacement rate or low normal stress, the response of the system is qualitatively different. Shear is distributed over the entire layer and flow velocity has a persistent, close-to-linear profile (Fig.~\ref{fig:RBC}b). A closer examination of shear strain rate, $\dot{\gamma} \equiv \partial{v_x}/\partial{y}$, reveals that $\dot{\gamma}$ exponentially decays from the walls to a constant value further away -- see the inset for the time-averaged shear rate. The excess shear rate near the boundaries is a non-local diffusive effect accompanying discontinuity in shear rate at the boundaries \cite{Koval2009,Bocquet2009,Kamrin2012,Kamrin2015,Pouliquen2009}. The effect of the excess shear rate on friction leads to a size dependence of $\mu$ \textit{vs.} $I$ relation, which will be analyzed in Sect.~\ref{sec:size}.

The center-of-mass velocity for the high inertial-number regime (Fig.~\ref{fig:RBC}d) fluctuates around $V/2$, which is the mean value of the time-averaged velocity profile shown in Fig.~\ref{fig:RBC}b by the dashed line. Fluctuations of velocity, both spatial and temporal, increase as the system approaches the transition to the shear-banding regime, $I \approx 10^{-3}$.

Friction in the distributed-shear regime is an increasing function of $I$ (Fig.~\ref{fig:RBC}e). The following phenomenological friction law (dashed line) is commonly used \cite{GDRMiDi,Forterre2008,daCruz2005} to capture the observed variation of friction over the entire studied range of $I \lesssim 10^{-1}$
\begin{equation}
\label{mu-I}
\mu (I) =\mu_\mathrm{qs} + a_\mu I^{b_\mu}  \,,
\end{equation}  
where $\mu_\mathrm{qs} = 0.258\pm0.002$, $a_\mu=0.85 \pm 0.05$, $b_\mu=0.81 \pm 0.04$ in agreement with previous studies \cite{daCruz2005,DeGiuli2017}; uncertainties correspond to the 95\% confidence interval estimated from the fitting procedure. The yield strength of the system, \textit{i.e.,} the minimum friction that permits continuous shear, is equal to $\mu_\mathrm{qs}$. In the shear-banding regime ($I<10^{-3}$) the friction coefficient is constant (neglecting a possible shallow minimum of friction) and equal to the yield strength, as the second term in Eq.~\ref{mu-I} is small. The second term becomes significant for $I>10^{-3}$, when the system transitions into the distributed-shear regime. The shear-banding regime is often called quasistatic, while the distributed-shear regime is termed inertial in the literature \cite{daCruz2005,GDRMiDi,Forterre2008,Jop,Koval2009}, reflecting the relative magnitudes of the inertial and stress forces. In this work we will use both variants interchangeably depending on whether our focus is on the level of strain localization or the magnitude of inertial number. In Sect.~\ref{sec:model} we will provide an explanation why the transition between the quasistatic and the inertial regimes coincides with the transition between the localized and the distributed shear regimes.

The friction coefficient under RBC, Eq.~\ref{mu-I}, is a sole function of the inertial number \added{and is independent of the contact stiffness number}. Indeed, Figure \ref{fig:RBC}e demonstrates that the $\mu(I)$ curve is independent of $N$, \textit{i.e.,} the whole dependence of friction on $N$ is through the dependence of $I(N)$. \deleted{In the present 2D system, $N$ is the ratio of normal stress and grain normal stiffness, $N=\hat{N}/\hat{k}_n$, where the asterisk denotes a dimensional quantity (in real units), \textit{e.g.} $\hat{N}$ in [Nm$^{-1}$] (in 2D) and $\hat{k}_n$ in [Nm$^{-1}$]. Analogously, in a three-dimensional (3D) system, $N$ would be defined as the ratio of normal stress and grain Young's modulus, $N=\hat{N}/\hat{E}$. This dimensionless number can be interpreted as the average compressive strain of a grain, $N=\hat{N}/\hat{k}_n \approx \delta/d$ because of the linear elasticity model employed in this work (see Sect.~\ref{sec:numerics}). Hence, $N$ represents softness of a grain (inverse stiffness): larger $N$ induce larger $\delta/d$, making grains effectively softer.} The observed independency of friction of \replaced{the contact stiffness number}{particle stiffness} under RBC \replaced{holds in the so-called rigid particle limit of $N\lesssim 10^{-3}$}{agrees with previous studies} \cite{deCoulomb2017,DeGiuli2017}\added{, which covers the range applied in this study}. \replaced{For higher $N$,}{Although} Refs. \cite{deCoulomb2017,Singh2015} reported a decrease of $\mu_\mathrm{qs}$ and porosity with $N$, and Ref. \cite{DeGiuli2017} reported an $N$-controlled transition from non-monotonic to monotonic $\mu(I)$ dependence. \deleted{, both these effects were activated only for softer grains ($N>10^{-3}$) than studied here.}

\section{Constant stress boundary conditions}

\subsection{Shear localization and friction law}\label{sec:SBC}

\begin{figure*}
\includegraphics{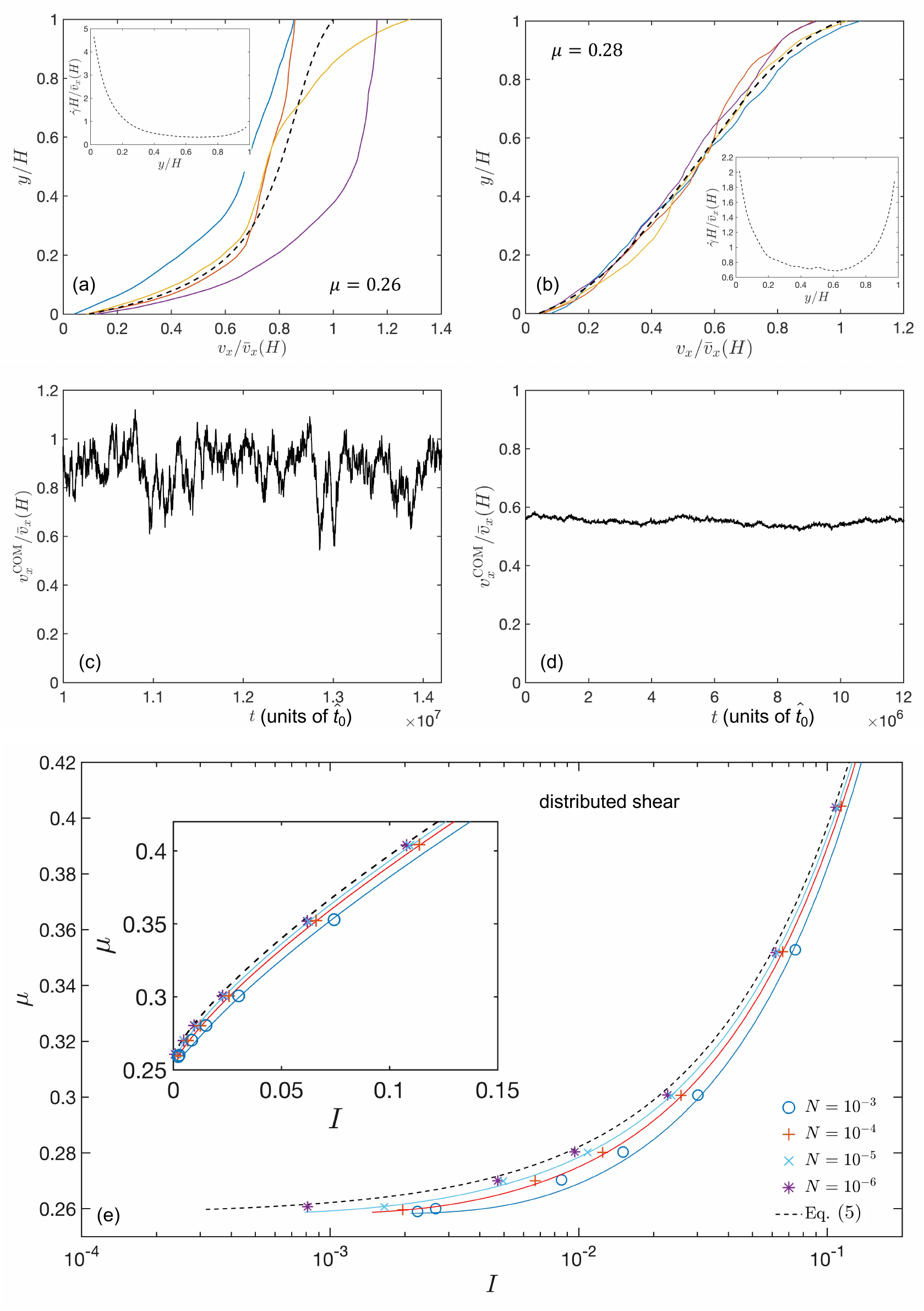}
\caption{Response of the layer under SBC. (a--b) Instantaneous shear-velocity profiles indicate distributed shear that is asymmetric (a) when $\mu-\mu_\mathrm{qs} < 10^{-2}$ and symmetric (b) for larger $\mu$. The dashed line denotes steady-state time average. Insets show shear strain rate deviating from a uniform value near boundaries. (c--d) Center-of-mass velocity as a function of time for the same conditions as in (a--b). (e) Friction coefficient \textit{vs.} inertial number. Symbols are numerical data and solid lines are fits to Eq.~(\ref{mu-I_SBC}). Different colors represent different values of normal stress $N$ as given in the legend.  The dashed line is the friction law under RBC, Eq.~(\ref{mu-I}), for reference. Inset shows the same data in a linear scale.}
\label{fig:SBC}
\end{figure*}

Unlike RBC, shear imposed by constant applied shear stress has been little studied so far. While RBC always generate continuous deformation, SBC may lead to transient deformation that ceases eventually, if driving shear stress is not strong enough. In Fig.~\ref{fig:SBC} we report results for samples and conditions that led to persistent shear. The left panel (a,c) displays results for a setup with $\tau=0.26 N$, $N=10^{-3}$ and $H=96$, \textit{i.e.,} friction coefficient $\mu=\tau/N$ is close to $\mu_\mathrm{qs}$, while the right panel (b,d) is for a setup with $\tau=0.28 N$, $N=10^{-6}$ and $H=96$, \textit{i.e.,} friction coefficient is well above the yield friction $\mu>\mu_\mathrm{qs}$.

In contrast to RBC, the shear-banding regime is absent and only distributed shear is observed, albeit the symmetric shear profile (Fig.~\ref{fig:SBC}b) becomes asymmetric (Fig.~\ref{fig:SBC}a) when the ratio of applied shear and normal stresses is close to the yield condition, $\tau/N - \mu_\mathrm{qs} < 10^{-2}$. Nevertheless, the asymmetric shear profile is more diffuse compared to shear bands observed under RBC, \textit{cf.} Fig.~\ref{fig:RBC}a. In addition, evolution of the center-of-mass velocity (Fig.~\ref{fig:SBC}c) shows smaller fluctuations compared to the intermittent dynamics observed in Fig.~\ref{fig:RBC}c for the shear-banding regime. Therefore, we reserve the term ``shear band" for a highly localized and intermittent shear state observed under RBC and low $I$, whereas the asymmetric and the symmetric shear states developed under SBC are both examples of the distributed-shear regime.

Figure \ref{fig:SBC}e shows variation of $I$ with $\mu$, in log-linear and linear scales. Note that while in the case of RBC we control the value of $I$ and the system responds with a value of $\mu$, under SBC we control the value of $\mu=\tau/N$ by setting the boundary stresses $\tau$ and $N$, and the system responds with a value of $I$. A layer with a given thickness $H$ under prescribed normal and shear stresses $N$ and $\tau$ thereby responds with shear strain rate $\dot{\gamma}$, resulting in a certain displacement rate $\bar{v}_x(H)$. Since any increase of $\tau$ leads to an increase of both $\mu$ (at a constant $N$) as well as the displacement rate, the rate-independent part of $\mu(I)$ is missing, \textit{cf.} Fig.~\ref{fig:RBC}e. 

Another interesting consequence of SBC is that $\mu$ is not a sole function of $I$, but, in addition, it systematically varies with the non-dimensional normal stress $N$ -- see Fig.~\ref{fig:SBC}e. Comparison with the friction law under RBC, Eq.~(\ref{mu-I}), indicated by the dashed line, shows that larger $N$ leads to a larger $I$ at a given $\mu$. \added{We propose the following mechanism to explain this}. \replaced{As discussed in Sect.~\ref{sec:units},}{Recall that} \deleted{the value of }$N$ is \replaced{the inverse of the contact stiffness number (Eq.~(\ref{kappa}))}{inversely proportional to grain stiffness}. 
\added{Larger $N$ therefore induces larger grain-level compressive strain $\hat{\delta}/\hat{d}$.} \replaced{Consequently, the elastic energy stored in compressed contacts, $\sim\hat{k}_n \hat{\delta}^2$, is an increasing function of $N$. The elastic energy builds up during collisions and configurational jamming, and afterward it is released in the form of kinetic energy, increasing grain inertia. Unless the collisions are fully dissipative, the released energy retains the increasing trend with $N$. Shearing at a fixed $\mu$, high $N$ conditions are therefore associated with larger elastic energy stored in contacts, which, when released, leads to larger grain inertia relative to low $N$ conditions. As a result, $I$ increases with $N$ at a fixed $\mu$, in agreement with Fig.~\ref{fig:SBC}e.}{
After overcoming a jammed configuration, the elastic energy stored in compressed contacts is released, increasing grain inertia. The released energy can be assumed to be a fraction of the total elastic energy $\hat{k}_n \hat{\delta}^2$. The elastic energy decreases with grain \added{normal} stiffness $\hat{k}_n$ at otherwise same conditions, because $\hat{\delta} \sim 1/\hat{k}_n$. Shear of soft grains at fixed $\mu$ is therefore associated with larger released energy and thereby larger grain inertia, \textit{i.e.} larger $I$, relative to that of rigid grains, in agreement with Fig.~\ref{fig:SBC}e.} 

The effect of \replaced{$N$}{grain softness} is much less pronounced for RBC, \textit{cf.} Fig.~\ref{fig:RBC}e. Since $\mu$ is not constrained under RBC, jammed configurations are associated with \replaced{increased shear force}{increase of $\mu$ relative to SBC, in agreement with Fig.~\ref{fig:fluct}}. The added shear force is used to overcome jammed configurations sufficiently fast to maintain the imposed boundary strain rate. This leads to increased dissipation and friction relative to SBC for the same $I$ (dashed line in Fig.~\ref{fig:SBC}e). It is likely that the increased friction can dissipate the elastic energy stored in compressed contacts, impeding the effect of $N$. The friction law under RBC, Eq.~(\ref{mu-I}), is therefore approached for low $N$ (in the hard-particle limit) under SBC, when the elastic energy is small. 

\deleted{We will first derive an empirical model to capture the observed variation with $N$, and we will discuss the mechanical origin of the effect of particle stiffness in Sect.~\ref{sec:stiffness}. }All data in Fig.~\ref{fig:SBC} were obtained for a single layer thickness $H=96$ to capture the $N$-dependence of the $\mu$ \textit{vs.} $I$ relation. In addition, there is also a size dependence for variable $H$, which will be discussed in Sect.~\ref{sec:size}.

The effect of $N$ can be \added{empirically} described as a shift in the value of $I$ relative to the value $I_0=\mu^{-1}(\mu)$ observed under RBC for given $\mu$: $\mu^{-1}(\mu)$ is an inverse function to Eq.~(\ref{mu-I}), which returns $I$ for given $\mu$. The shift $I-I_0$ of the inertial number for SBC relative to that for RBC is found to depend on $N$ as $\sim N^{0.3}$ at fixed $\mu$. In addition, it also increases with the proximity to the yield friction coefficient as $\sim I_0/(\mu-\mu_\mathrm{qs})$. This leads to $I-I_0=c_N N^{0.3}I_0/(\mu-\mu_\mathrm{qs})$. Consequently, the friction law under SBC can be expressed as 
\begin{equation}
\label{mu-I_SBC}
I=\mu^{-1}(\mu)\left[1+ c_N \frac{N^{0.3}}{\mu-\mu_\mathrm{qs}}\right] \approx \left(\frac{\mu-\mu_\mathrm{qs}}{0.85}\right)^\frac{1}{0.81} \left[1+ c_N \frac{N^{0.3}}{\mu-\mu_\mathrm{qs}}\right] 
\end{equation}  
with $c_N=0.17 \pm 0.03$ obtained from a fit to the simulation data. Equation (\ref{mu-I_SBC}) for various $N$ is depicted in Fig.~\ref{fig:SBC}e (solid lines). 

The dimensional variant with dimensional $\hat{N}$ [Pa] (in 3D) reads
\begin{equation}
\label{mu-I_SBC_dim}
I=\mu^{-1}(\mu)\left[1+ c_N \frac{(\hat{N}/\hat{E})^{0.3}}{\mu-\mu_\mathrm{qs}}\right] \,,
\end{equation}  
where $\hat{E}$ [Pa] is Young's modulus of grains. 

\begin{figure}
\includegraphics[width=0.5\textwidth]{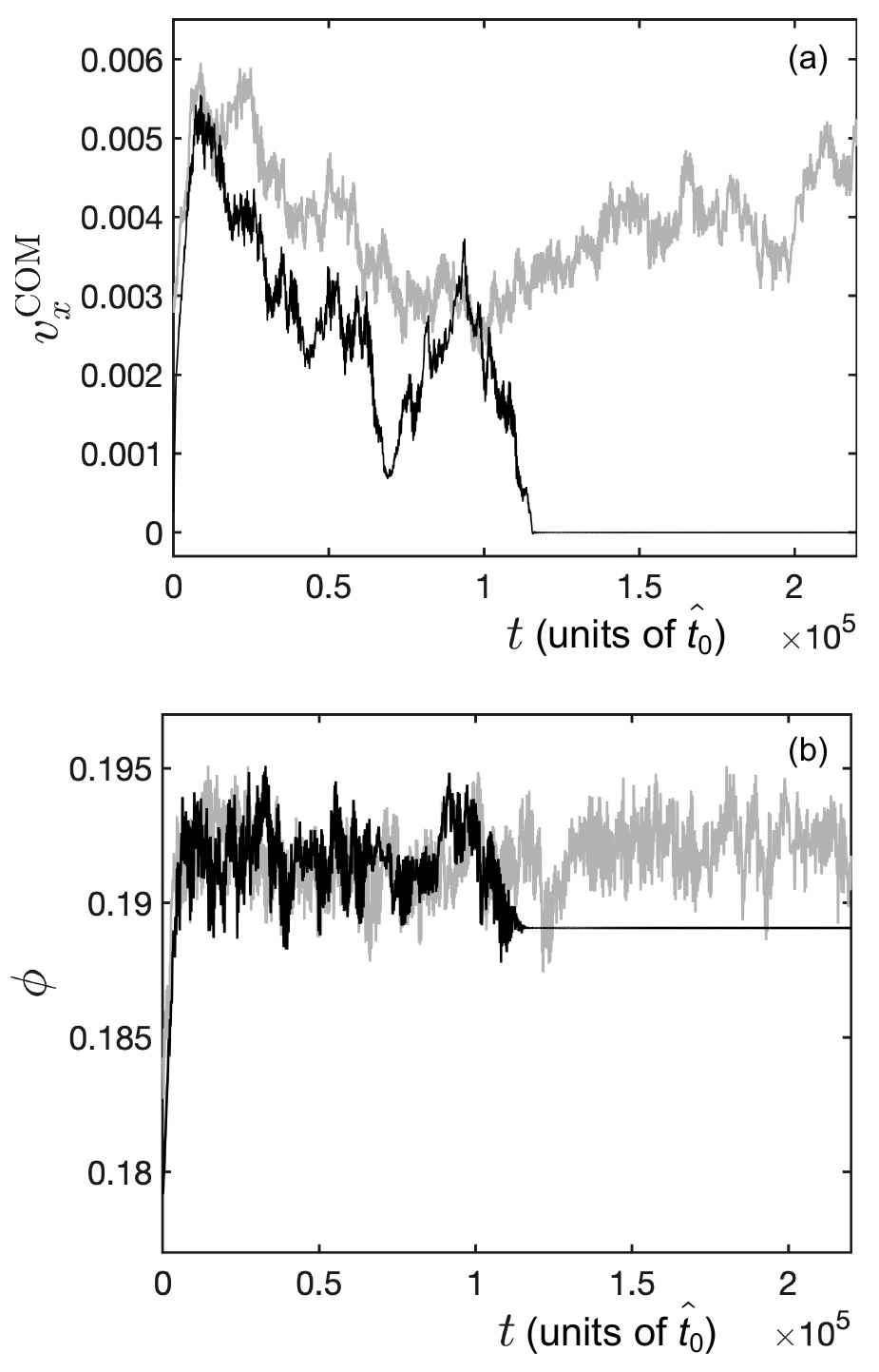}
\caption{Evolution of center-of-mass velocity (a) and porosity (b) for two distinct samples (differing in positions and sizes of grains, although randomly drawn from the same distribution) under SBC for $N=10^{-3}$, $\tau=0.26 N$ ($\mu=0.26$) and $H=48$. Despite the similarity in initial evolution, one sample (light line) reaches steady flow, while the other (dark line) stops after a transient deformation.}
\label{fig:eff_ini}
\end{figure}

Yield strength observed under SBC tends to the same value as for RBC, $\mu_\mathrm{qs} = 0.258$. Under lower applied shear stress, $\tau<\mu_\mathrm{qs} N$, only transient deformation occurs and the system eventually stops. However, not all systems at $\tau/N \geq \mu_\mathrm{qs}$ reached steady flow. The transient dynamics is found sensitive to sample preparation. Some samples deformed transiently and stopped even for $\tau/N$ up to $\approx 0.28$, while other samples for the same conditions did flow steadily. A comparison of time series of two different initial configurations for otherwise same conditions ($N=10^{-3}$, $\tau=0.26 N$ and $H=48$) is shown in Fig.~\ref{fig:eff_ini}. Despite the similarity of initial evolutions of center-of-mass velocity and porosity \deleted{(defined as volumetric fraction of voids)}, $\phi$, one sample (light line) reached steady flow while the other (dark line) stopped after a transient deformation. We therefore conclude that the yield strength under SBC is sample-dependent and thereby it is not, strictly speaking, a proper material parameter as for RBC. \added{Porosity, $\phi$, in this manuscript is conventionally defined as the volume fraction of voids.}\footnote{Porosity $\phi$ should not be confused with solid (volume) fraction, equal to one minus porosity, which is unfortunately also often denoted by the same symbol.}

\added{The transient flow for both SBC and the inertial regime of RBC is free of shear bands. Slip is initiated along a weak plane in the material, typically near one of the boundaries. Shear rate spreads out from the initial slip zone via a diffusion front that controls the transient time \cite{Shojaaee2012a,Parez2016}. The transient phase is proportional to the time it takes for the diffusion front to swipe through the system. For planar shear, the transient time is proportional to the square of layer thickness and inversely proportional to the normal stress and a rate-strengthening coefficient, $\mathrm d \mu/ \mathrm d \hat{\dot{\gamma}}$. On the other hand, time to stop for a system under SBC in the vicinity of yield stress is a random variable depending on how long the system wonders in its phase space before it reaches a mechanically stable microscopic configuration \cite{Clark2018}. As a result, the system may jam earlier or later than flow becomes fully developed.
}

The lack of steady flow under SBC when $\mu \to \mu_\mathrm{qs}$ is related to the fact that the force exerted on the system is fixed by \replaced{$\tau L^{D-1}$}{$\tau L$} in the case of SBC, whereas it is unbound in the case of RBC. Consequently, when the system hits a strong asperity\footnote{In DEM an asperity may be thought of as a grain configuration that requires above-average shear stress for deformation to proceed, either by dilatancy or by enhanced dissipation.}, the force might not be sufficient to break it under SBC and the system eventually stops. This hypothesis is confirmed in Fig.~\ref{fig:fluct}. Here we examine the effect of boundary conditions on fluctuations of shear stress measured at the moving boundary, $\mathrm{std}(\sigma_{xy}(H))$, where $\mathrm{std}(\cdot)$ denotes the standard deviation of a time series. The boundary shear stress $\sigma_{xy}(H)$ is calculated as the x-component of the force acting on the boundary due to contacts with internal grains, normalized by the length of the boundary (area in 3D), \replaced{$\sigma_{xy}(H)=F_x/L^{D-1}$}{$\sigma_{xy}(H)=F_x/L$}. Note that $\sigma_{xy}$ fluctuates even for SBC because of fluctuations of contact forces with grains adjacent to the wall. The fluctuations of $\sigma_{xy}$ are a measure of the range of shear forces experienced by the layer in the course of deformation: the largest magnitude forces arise to overcome strong asperities (jammed states), while the smallest magnitude forces occur as the system unjams and accelerates. A similar picture was introduced in Ref.~\cite{Muhlhaus1987}, where fluctuations of shear stress were attributed to the buildup and collapse of force chains.

Figure \ref{fig:fluct} demonstrates that RBC (open symbols) is indeed able to generate larger stresses to overcome asperities. The stress fluctuations grow as $\sim I^{0.7}$ in the inertial regime, as a result of increasingly stronger collisions. However, the coefficient is lower in the case of SBC relative to RBC. The SBC data suggest that deformation ceases if stress fluctuations fall below about $10^{-2}$. This happens at $I \simeq 10^{-3}$. Lower fluctuations of shear stress are not sufficient to overcome fluctuations in strength that arise during shear. For RBC, on the other hand, the standard deviation of shear-stress fluctuations saturates in the quasistatic regime at $\approx 2\cdot 10^{-2}$. The larger magnitude of fluctuations for RBC means that shear stress generated during a jamming episode momentarily exceeds shear stress encountered during jamming under SBC. Therefore RBC runs are able to overcome and move past configurations that will cause SBC runs to stop. \deleted{The normalized stress fluctuations shown in Fig.~\ref{fig:fluct} also depend on $N$ in the inertial regime; this effect will be discussed in Sect.~\ref{sec:stiffness}.}

\begin{figure}
\includegraphics[width=0.5\textwidth]{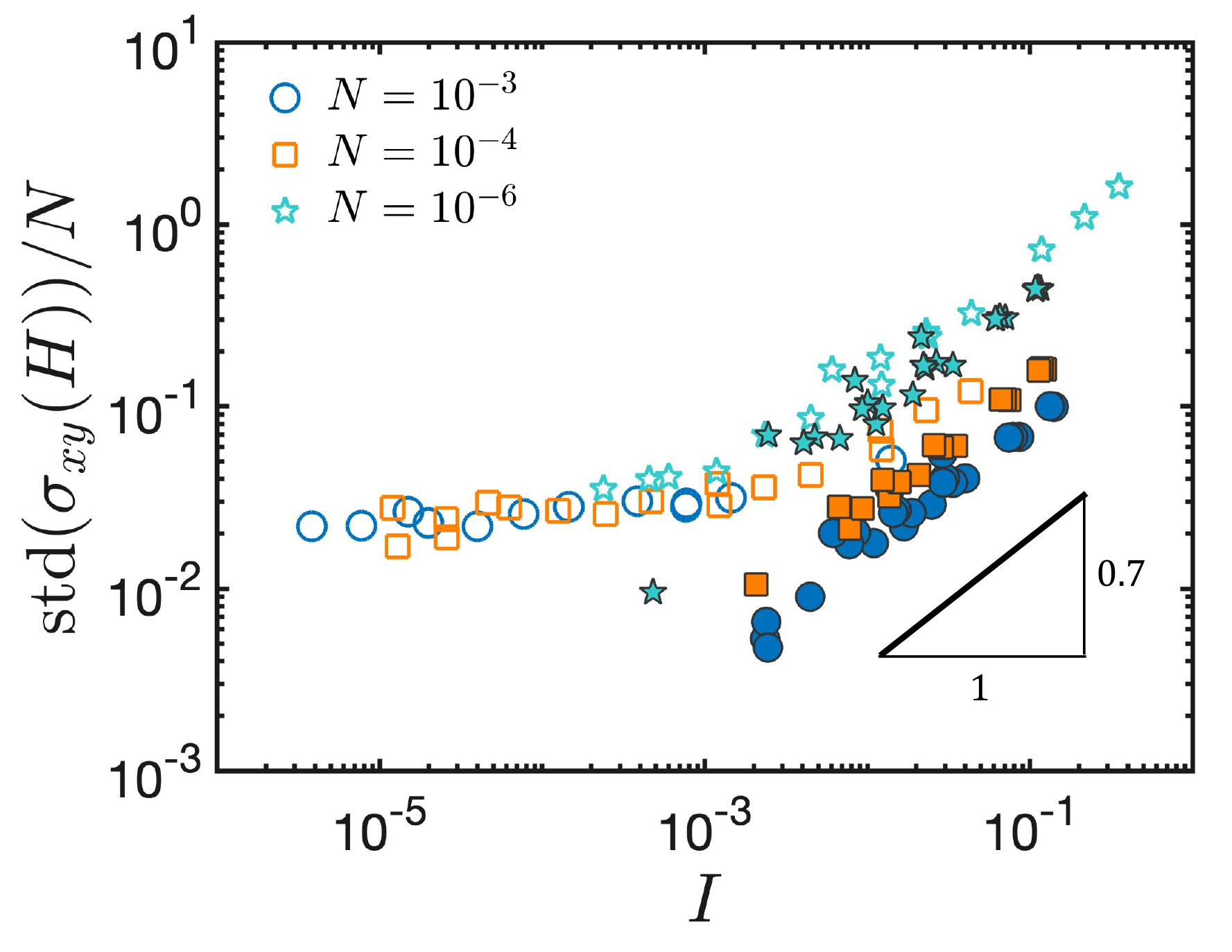}	
\caption{Effect of boundary conditions on fluctuations of shear stress, normalized by the applied normal stress. The fluctuations, calculated as the standard deviation of the boundary shear stress, reflect the magnitude of above-average shear stress that is momentarily generated to overcome strong force chains or asperities. The stress fluctuations are larger for RBC (open symbols) than for SBC (filled symbols), pointing at the ability of RBC to sustain steady deformation for any $I$. Deformation under SBC ceases for $I \lesssim 10^{-3}$ once the fluctuations drop sufficiently below their quasistatic limit $\approx 2\cdot 10^{-2}$. Reduced shear-stress fluctuations imply restricted ability to overcome fluctuations in strength of the shearing layer. Different symbols and colors represent different values of normal stress as given in the legend.}
\label{fig:fluct}
\end{figure}

\added{In the framework of constitutive modeling, non-local models allow interpretation of shear localization as an instability by which local plastic deformation induces redistribution of the elastic stress within a predicted ``cooperativity distance," potentially triggering other plastic events \cite{Bocquet2009,Kamrin2012}.} \replaced{A similar argument was used by DeGiuli and Wyart \cite{DeGiuli2017}, who identified the key dynamic parameter controlling local shear rate with acoustic noise}{DeGiuli and Wyart \cite{DeGiuli2017} hypothesized that shear localization into shear bands is associated with an instability by which collisions between grains produce acoustic noise that helps trigger slip on near, locked contacts, generating more collisions}. \added{Whether the flow law is controlled by rate of plastic events or by acoustic noise,} fluctuations of shear stress in Fig.~\ref{fig:fluct} can be viewed as a measure of such mechanical noise. RBC always generates sufficient noise to sustain shear of the layer owing to the unlimited energy input from the boundaries. In contrast, limited energy input under SBC generates sufficient noise only for rapid enough sliding. This is manifested by larger $I$ required under SBC to generate the same magnitude of fluctuations as for RBC. In addition, fluctuations of shear stress decrease with increasing $N$ \added{at constant $I$} in a similar way \replaced{found for acoustic noise}{that mechanical noise decreases for softer grains (larger $N$), as found in} \cite{DeGiuli2017}.

\subsection{Dilatancy law}
\label{sec:dilatancy}

\begin{figure}
\includegraphics[width=0.5\textwidth]{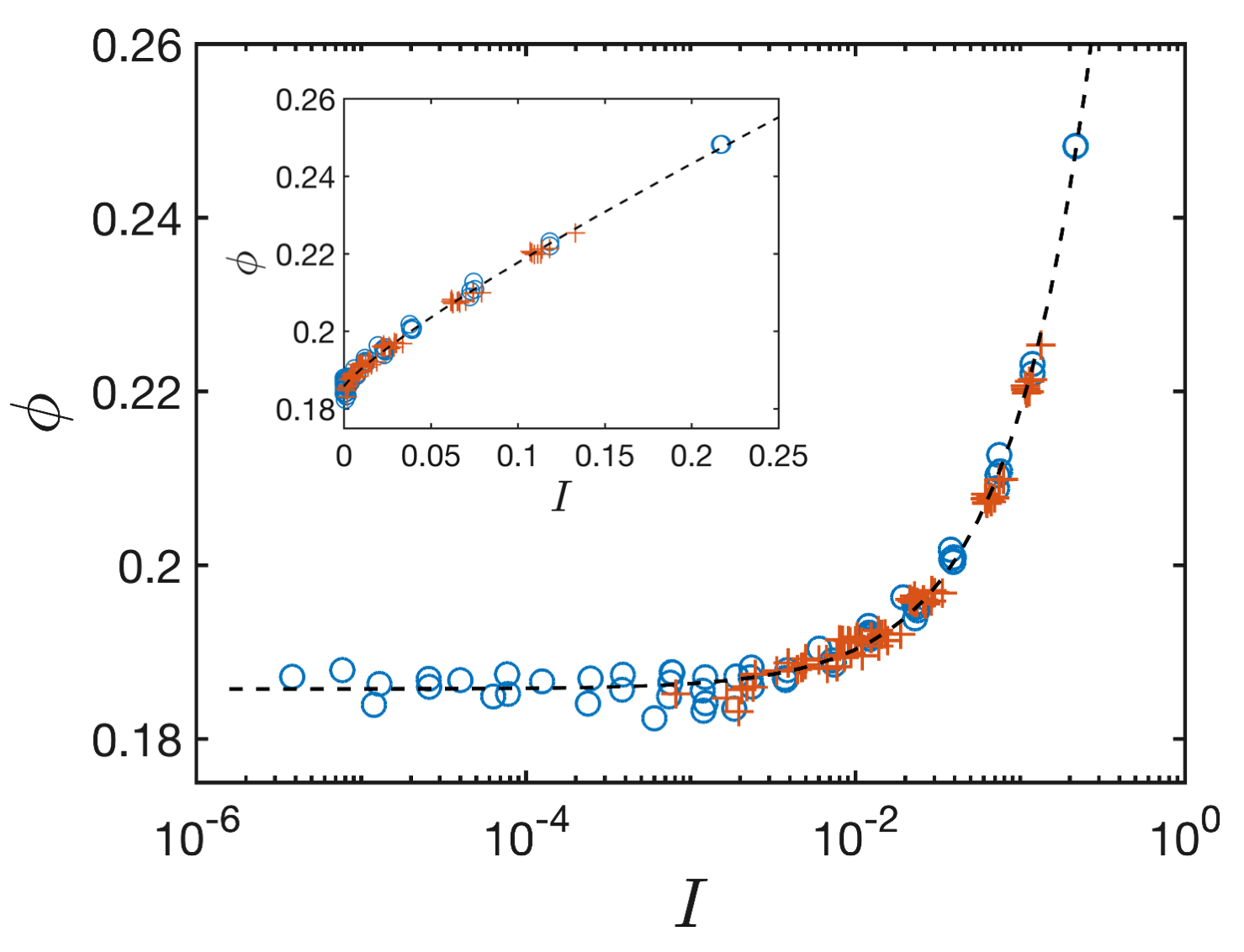}	
\caption{Porosity is controlled by inertial number. The same functional dependence is observed for RBC (circles) as well as SBC (`+' symbols). The dashed line is a fit to Eq.~(\ref{phi-I}). The inset displays the same data in a linear scale.}
\label{fig:phi}
\end{figure}

Similarly to friction, porosity of sheared granular media under RBC is controlled by inertial number as expressed by a function $\phi(I)$, referred to as the dilatancy law \cite{daCruz2005}. Figure \ref{fig:phi} demonstrates that the same function characterizes porosity also under SBC. Both RBC and SBC data under various conditions collapse onto a single master curve that is well described by a similar functional form as for to the friction law
\begin{equation}
\label{phi-I}
\phi (I) =\phi_\mathrm{qs} + a_\phi I^{b_\phi}  \,,
\end{equation}  
where $\phi_\mathrm{qs}  = 0.1858 \pm 0.0004$, $a_\phi=0.224 \pm 0.004$, $b_\phi=0.85 \pm 0.02$ with uncertainties corresponding to the 95\% confidence interval estimated from the fitting procedure. 

Data for SBC fall within the inertial regime, $I \geq 10^{-3}$, as already discussed above. No dependence of porosity on \replaced{the contact stiffness number}{stiffness} is observed, contrary to the case of friction coefficient under SBC. In other words, $\phi$ is a sole function of $I$ and no additional dependence on $N$, besides the intrinsic dependence $I(N)$, arises. \deleted{Recall that the dimensionless normal stress $N$ can be interpreted as the average compressive strain of a grain. }The applied normal stress levels induce grain compressive strain of $\delta/d=10^{-7} - 10^{-3}$. Such strain has a negligible effect on porosity \replaced{in the inertial regime, \textit{cf.}}{compared to the inertial effect shown in} Fig.~\ref{fig:phi}. While the effect of \replaced{the contact stiffness}{finite stiffness} leads to compression of the system's volume by a factor no higher than $10^{-3}$, the inertial effect induces expansion by a factor $\geq 10^{-2}$ relative to the volume in the quasistatic regime, corresponding to $\phi_\mathrm{qs}$. Inertial effects thus dominate the value of porosity in the inertial regime, which explains the observed independency of the \replaced{$\phi(I)$ curve}{porosity} of $N$. In the quasistatic regime, decrease of porosity \added{$\phi_\mathrm{qs}$} with $N$ was reported in \cite{deCoulomb2017,Singh2015}, but only \replaced{for}{in the soft particle regime of} $N>10^3$\added{, \textit{i.e.,} higher normal stresses than applied here}.

\deleted{Note that the range of applied normal stress reflects realistic values for the geological setting considered. Using elastic modulus $5\cdot 10^{10}$ Pa of rocks (\textit{e.g.}, granite), normal stress levels considered here range $\hat{N}\approx 10^4-10^8$ Pa, which is typical for landslides and (not too deep) natural faults, \textit{e.g.} \cite{Smeraglia2017,Marone1990,LOGAN1992}.}

\begin{figure}
\includegraphics[width=0.5\textwidth]{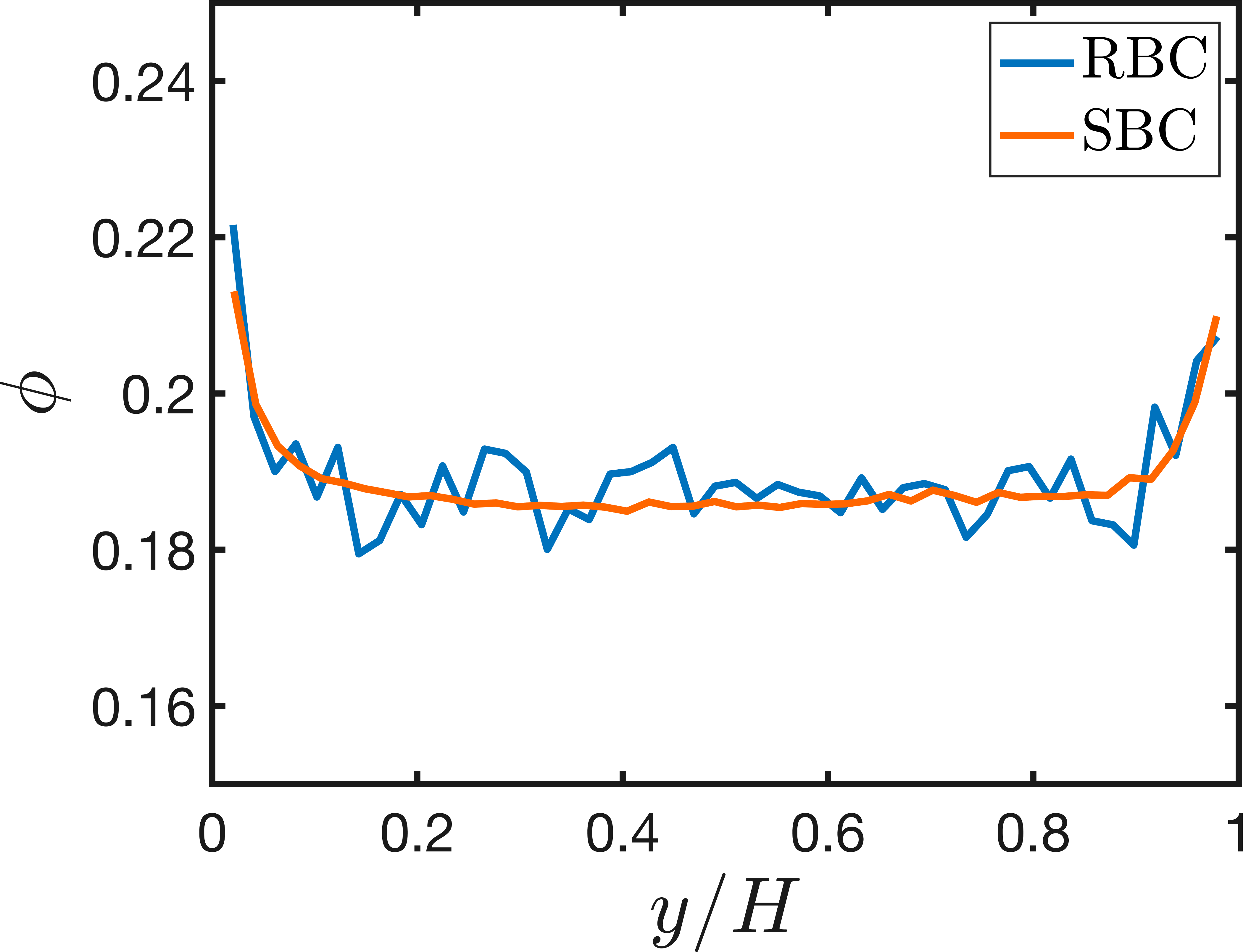}	
\caption{Porosity profiles for RBC (blue) at $I=4 \cdot 10^{-5}$, $N=10^{-3}$ and $H=96$ (same setup as in Fig.~\ref{fig:RBC}a, c) and SBC (orange) at $\mu=0.26$, $N=10^{-3}$ and $H=96$ (same setup as in Fig.~\ref{fig:SBC}a, c). Both setups lead to overall porosity $\phi=0.1864\pm0.0004$, close to the minimum value $\phi_\mathrm{qs}$.}
\label{fig:phi-y}
\end{figure}

Porosity in the present 2D system is calculated as the area fraction of voids in the total area of the layer excluding boundary regions of thickness 6 grains near each wall. Within the boundary regions porosity increases toward the wall due to the excluded-volume effect, as shown in Fig.~\ref{fig:phi-y}. Further away from the walls, porosity is uniform.

A comparison of RBC and SBC runs reveals similarity of the time-averaged porosity profiles (Fig.~\ref{fig:phi-y}). The selected profiles were obtained for conditions resulting in the overall porosity being close to the minimum value $\phi_\mathrm{qs}$, for both the RBC and the SBC runs. The scatter in local porosity is notably larger in the RBC case, despite the number of independent time frames used in the calculation of the averaged profiles is the same for the two boundary conditions. The larger fluctuations of porosity under RBC are associated with porosity changes between periods when the given point is within a randomly migrating shear band (high porosity) and when it is in the spectator static or creeping region (low porosity). This hypothesis will be further elaborated in Sect.~\ref{sec:model} (Fig.~\ref{fig:cartoon}). 

\subsection{Role of layer thickness}
\label{sec:size}
\begin{figure}
\includegraphics[width=0.5\textwidth]{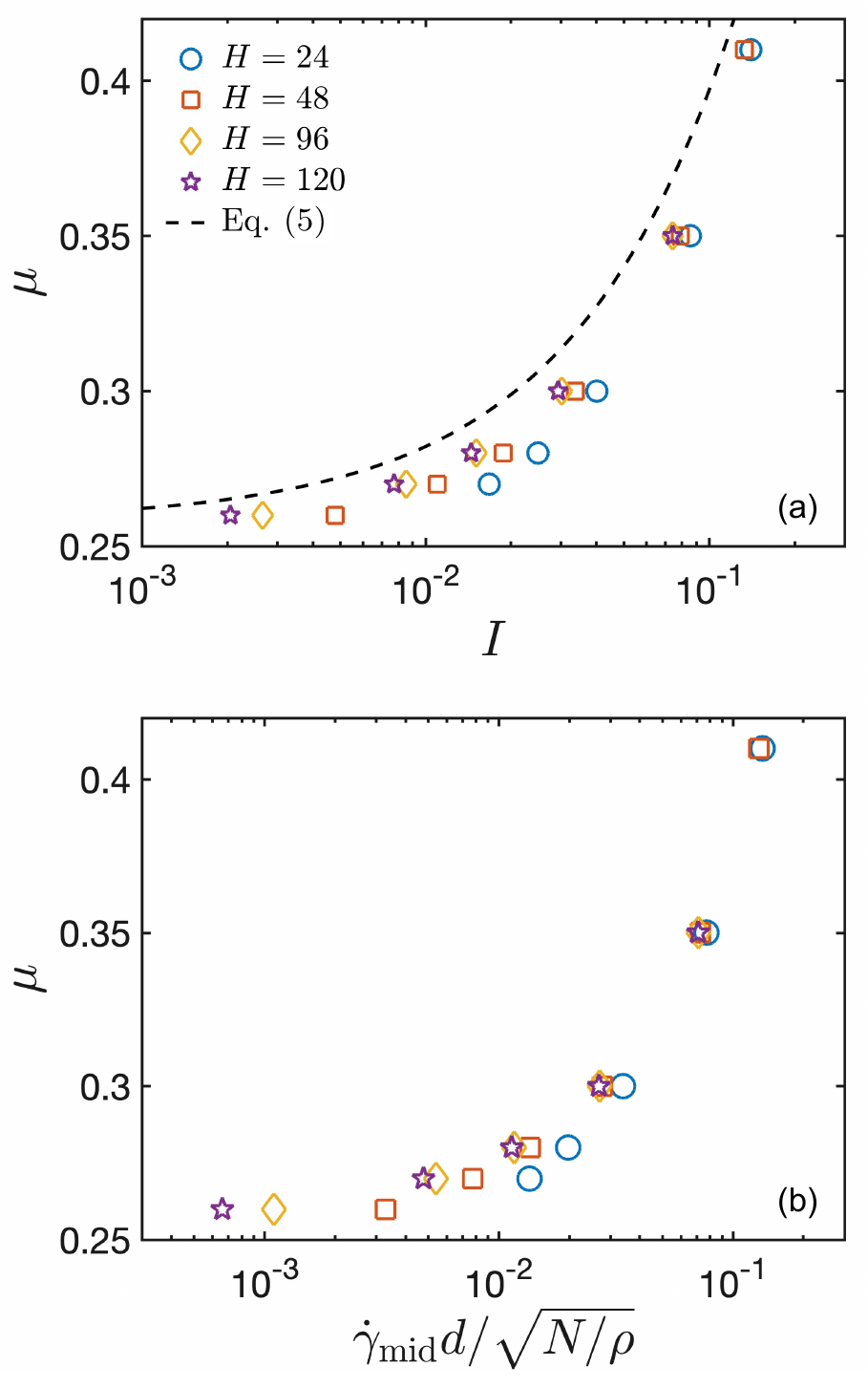}
\caption{Effect of layer thickness $H$ on the friction law under SBC for $N=10^{-3}$ and various $H$ as given in the legend. (a) $\mu$ \textit{vs.} $I$. The dashed line is Eq.~(\ref{mu-I}) representing the friction law under RBC. (b) $\mu$ \textit{vs.} $I_\mathrm{loc}$, \textit{i.e.,} the local inertial number in the center of the layer. $I_\mathrm{loc}$ is independent of $H$ for fixed $\mu$ and $N$ provided that $H$ is large enough to support uniform shear-rate region around the center. The last condition is violated when $\mu$ is sufficiently close to $\mu_\mathrm{qs}=0.258$ or when $H$ is small (blue circles).}
\label{fig:size}
\end{figure}

It should be noted that both friction and dilatancy laws, $\mu(I)$ and $\phi(I)$, plotted in Figs.~\ref{fig:RBC}e, \ref{fig:SBC}e and \ref{fig:phi}, suffer from slight dependence on layer thickness $H$. \added{Figure \ref{fig:size}a shows the effect of $H$ on the friction coefficient for $N=10^{-3}$ and SBC.} \deleted{We observe this boundary effect only in the distributed shear regime. }The $H$ dependence arises because the mean shear rate $v_x(H)/H$, which enters $I$, is affected by boundary contributions due to tails displayed in the insets of Figs.~\ref{fig:RBC}b and \ref{fig:SBC}a--b. These tails are diffusive features accompanying discontinuity in shear strain rate at the boundaries \cite{Koval2009,Kamrin2012,Kamrin2015,Pouliquen2009}.  

\added{The excess shear rate due to the presence of boundaries, $\dot{\gamma}_\mathrm{e}$, decays exponentially with the distance from the boundary.} The width, \textit{i.e.,} the decay length, of the exponentially decaying tails is a function of friction coefficient $\sim 1/(\mu-\mu_\mathrm{qs})^\alpha$ with $\alpha = 0.4-0.5$ consistent with Refs. \cite{Kamrin2012,Kamrin2015}. \added{The $\mu$-dependent width makes it impossible to fully remove the boundary effects with increasing thickness, because the tails eventually span the entire layer if $\mu$ approaches $\mu_\mathrm{qs}$. To estimate the relative magnitude of the boundary effects, we write the total shear rate as a sum of the excess shear rate and the bulk shear rate, $\dot{\gamma}_\mathrm{b}$: $\dot{\gamma}=\dot{\gamma}_\mathrm{e}+\dot{\gamma}_\mathrm{b}$. The bulk shear rate is the shear rate in the homogeneous shear region, which occurs around the center of the layer (far away from the boundaries) in the limit $H \to \infty$ (at a fixed $\mu$). The average shear rate, which appears in the definition of $I$, is $\bar{v}_x(H) / H = (1/H) \int (\dot{\gamma}_\mathrm{b} + \dot{\gamma}_\mathrm{e}) \mathrm dy = \dot{\gamma}_\mathrm{b} + (1/H)\int \dot{\gamma}_\mathrm{e} \mathrm dy$. The observed $H$ dependence is due to the last term, which decreases with increasing $H$ at a fixed $\mu$, but increases with decreasing $\mu-\mu_\mathrm{qs}$ at a fixed $H$. In other words, the $H$ dependence vanishes for thick enough layers, for which the bulk region is much larger than the boundary region, but this condition requires increasingly larger thickness when approaching the quasistatic conditions ($\mu \to \mu_\mathrm{qs}$), in agreement with Fig.~\ref{fig:size}a.} \deleted{Decomposing the total shear rate into the excess shear rate (due to the boundary effects), $\dot{\gamma}_\mathrm{e}$, and the bulk shear rate in the homogeneous region, $\dot{\gamma}_\mathrm{b}$: $\dot{\gamma}=\dot{\gamma}_\mathrm{e}+\dot{\gamma}_\mathrm{b}$, and evaluating the inertial number $I \sim \bar{v}_x(H) / H = (1/H) \int \dot{\gamma}_\mathrm{e}+\dot{\gamma}_\mathrm{b} \mathrm dy$ (\textit{cf.} Eq.~(\ref{I})), it is clear that the contribution of the first righ-hand term, $(1/H) \int \dot{\gamma}_\mathrm{e}\mathrm dy$, increases with decreasing $H$, since $\dot{\gamma}_\mathrm{e}$ is independent of $H$ and quickly decays to 0 away from the boundaries. In other words, boundary tails occupy a fixed space, but this space is a larger portion of the layer for smaller $H$. As a result, $H$ has a negative feedback on $I$ for a given $\mu$.}

\deleted{The width of the boundary tails increases when approaching the yield condition, so that the tails eventually span the entire system unless a transition into the shear-banding regime occurs (for RBC). Consequently, the effect of $H$ on $I$ is stronger the closer $\mu$ is to $\mu_\mathrm{qs}$.}

\replaced{Figure \ref{fig:size}b demonstrates that the size effect cannot be removed by considering a local inertial number $I_\mathrm{loc}=\hat{\dot{\gamma}}_\mathrm{mid}\hat{d}/\sqrt{\hat{N}/\hat{\rho}}$, defined using the local shear strain rate in the center of the layer, $\dot{\gamma}_\mathrm{mid}=\dot{\gamma}(H/2)$. Although the collapse of the data for different $H$ for $\mu>0.3$ is better than in Fig.~\ref{fig:size}a, a significant $H$ dependence persists for lower $\mu$. The reason is lack of the uniform-shear, bulk region in the center of the layer under the latter conditions. As $\mu$ approaches $\mu_\mathrm{qs}$, the boundary tails penetrate into the center of the layer, and $\dot{\gamma}_\mathrm{mid}$ contains a contribution from $\dot{\gamma}_\mathrm{e} \sim \mathrm e^{-H/2\xi}$, where $\xi \sim 1/(\mu-\mu_\mathrm{qs})^\alpha$ is the decay length \cite{Kamrin2012,Kamrin2015}.}{
In Fig.~\ref{fig:size}a we plot $\mu$ \textit{vs.} $I$ relation under SBC for $N=10^{-3}$ and variable $H$ as given in the legend. Indeed, $\mu(I)$ curves for smaller $H$ are shifted to the right (larger $I$). The effect of $H$ becomes prominent as $\mu$ approaches $\mu_\mathrm{qs}$. Since the $H$ dependence of $\mu(I)$ curves is due to the boundary effects, it could be removed by using the inertial number defined as $I=\hat{\dot{\gamma}}_\mathrm{b}\hat{d}/\sqrt{\hat{N}/\hat{\rho}}$, which would be measured far away from the boundaries, where shear rate is uniform ($\dot{\gamma}=\dot{\gamma}_\mathrm{b}$). Figure \ref{fig:size}b shows the dependence of $\mu$ on the local inertial number in the center of the layer, $I_\mathrm{loc}=\dot{\gamma}_\mathrm{mid}d/\sqrt{N/\rho}$, where $\dot{\gamma}_\mathrm{mid}=\dot{\gamma}(H/2)$ serves as a proxy for $\dot{\gamma}_\mathrm{b}$. However, using $\dot{\gamma}_\mathrm{mid}$ only marginally removes the size dependence. Data for various $H$ are identical as long as $H$ is large enough to support uniform shear rate region around the center of the layer, \textit{e.g.} $H>24$ and $\mu>0.3$. As $\mu$ approaches $\mu_\mathrm{qs}$, the boundary tails eventually penetrate the center of the layer and $\dot{\gamma}_\mathrm{mid}$ becomes $H$-dependent because of the contribution from $\dot{\gamma}_\mathrm{e}$. Since $\dot{\gamma}_\mathrm{e}$ in the center of the layer is proportional to $\mathrm e^{-H/2\xi}$~\cite{Kamrin2012,Kamrin2015}, where $\xi$ is the decay length, its effect becomes significant once $\xi(\mu) \sim H$.}

The $H$ dependence of $\mu(I)$ curves, shown in Fig.~\ref{fig:size} for SBC, also exists for RBC because the excess shear rate is roughly independent of the boundary conditions. However, the effect of $H$ is mitigated for RBC near $\mu \approx \mu_\mathrm{qs}$, where it is the strongest for SBC, because of the transition to the shear-banding regime. This is evident from low scatter of data in Fig.~\ref{fig:RBC}e, calculated for a range of $H=24-200$, compared to Fig.~\ref{fig:size}a. Note also that data in Fig.~\ref{fig:size}a are shifted to the right to the friction law under RBC, Eq.~(\ref{mu-I}), represented by the dashed line, even for large $H$. This is due to the $N$ dependence whereby $I$ increases with $N$ for fixed $\mu$ and $H$, as described by Eq.~(\ref{mu-I_SBC}).

\added{The effect of $H$ is likely to be mitigated for rougher boundaries, because rough walls hamper generation of excess shear. The effect of roughness was studied \textit{e.g.,} in Ref.~\cite{Shojaaee2012b}.}

\section{Connection between porosity variation and strain localization}\label{sec:model}
In this section, we introduce a simple model for shear localization based on a) critical porosity required for shear, similar to the concept of the critical state in soil mechanics \cite{Lambe1969}, and b) relation between porosity changes and friction, developed in \cite{Frank1965,Marone1990}. 

Dilatancy plays a critical role in accommodating shear in granular media, so that compacted material needs to dilate for shear to commence, as already realized by Reynolds \cite{Reynolds1885}. On the other hand, dilatancy incurs a penalty on the energy budget due to work performed against the applied normal stress. This energy penalty provides a direct, causal relation between dilation and friction, which has been demonstrated in geophysical experiments that study friction in geological faults \cite{Marone1990,Marone1991} and has also been studied theoretically \cite{Rowe1962,Frank1965}. The theoretical analysis, relating friction and dilation, is next reviewed and then applied to explain localization in the quasistatic regime and absence of localization in the inertial regime. \deleted{For clarity, we switch to dimensional quantities in the rest of the manuscript and drop the asterisks superscript $^*$ introduced earlier.}

Work expended during shear per unit volume of material is $\tau \mathrm d \gamma$, where $\mathrm d \gamma$ is shear strain increment. This work can be decomposed into work done against contact forces, $\tau_f \mathrm d \gamma$, where $\tau_f$ is shear stress on slipping granular contacts, and work done against the normal stress, $N \mathrm d \epsilon_V$, where  $\mathrm d \epsilon_V = \mathrm d H/H =  \mathrm d \phi/(1-\phi)$ is volumetric-strain increment \cite{Frank1965,Marone1990}
\begin{equation}
\label{w}
\tau \mathrm d \gamma=\tau_f \mathrm d \gamma+ N \mathrm d \epsilon_V \,.
\end{equation}  
The first term on the right-hand side is associated with dissipation in granular contacts, while the second term is work associated with volume change of the layer. 
The last equation can be rewritten in terms of friction, dividing it by $ N\mathrm d \gamma$,
\begin{equation}
\label{mu-dH}
\mu=\mu_f + \frac{\mathrm d \epsilon_V}{\mathrm d \gamma}  \,,
\end{equation}  
where $\mu_f = \tau_f / N$.

Equation (\ref{w}) constrains microscopic configurations of grains to those for which the dissipation due to displacements $\mathrm d \gamma$ and $\mathrm d \epsilon_V$ does not exceed the amount of energy supplied by the boundaries, $\tau \mathrm d \gamma$. In the lowest-energy shear state, the layer dilates only locally within a shear band to minimize the last term associated with an increase in global volumetric strain $\mathrm d \epsilon_V$. Indeed, if dilatancy is limited to the $\Delta y$ required to overcome asperities on a single (weakest) plane, then the increment of global volumetric strain is only $\Delta y / H$, albeit local volumetric strain over a shear band of thickness $h<<H$, $\Delta y / h$, is much larger. In contrast, for uniform shear the global and local volumetric strains are the same and of the order of $\Delta y / h$. Therefore, uniform shear requires larger volumetric change than localized shear. 

When the work expended in the system is increased, \textit{e.g.,} as in the inertial regime, kinetic energy due to grain collisions provides configurations with larger $\mathrm d \epsilon_V$. In other words, under higher-energy conditions the mean porosity increases, reducing the contrast between the porosity inside a shear band and that of the background. Once the global porosity is uniformly equal to (or exceeds) the critical porosity originally present only within the shear band, there is no longer incentive for localization, and the layer shears with uniform shear rate.

This idea agrees with the quasistatic shear experiments of \cite{Desrues1996}, which show that initially loosely packed sand layers have no incentive to localize, while initially densely packed layers localize by increasing porosity to the critical porosity value only within a shear band. The experiments of \cite{Marone1990} show that dense granular layers transiently deform via distributed shear, but steady-state shear is accommodated by localized shear structures that form as soon as the peak stress is achieved during constant-rate loading. However, the previous works did not investigate the connection of localization to the stress and rate conditions. 

In an attempt to explain the transition from localized to distributed shear as the inertial number increases, we suggest a simple mechanical model based on the following assumptions: (a) shear, whether localized or distributed, requires local porosity to attain a critical value, $\phi(y) \geq \phi_\mathrm c$, and (b) global porosity is dictated by inertial number, as in Fig.~\ref{fig:phi}, reflecting the growing agitation of the system due to inertial effects.

\begin{figure}
\includegraphics[width=0.5\textwidth]{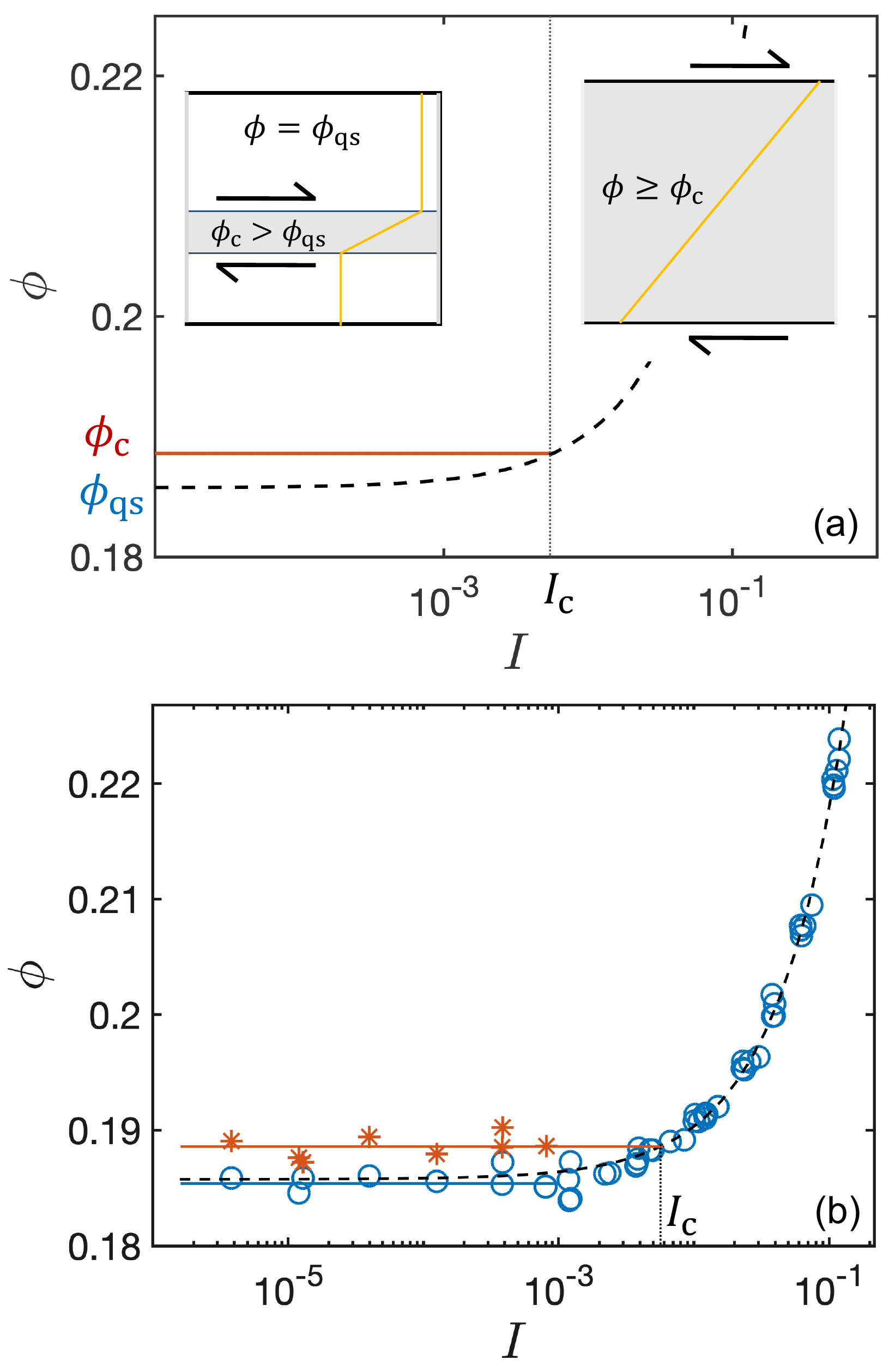}	
\caption{Transition from localized to distributed shear is facilitated by enhanced porosity due to inertial effects. (a) Schematic of the transition. In the quasistatic regime of small $I$, porosity attains the critical value $\phi_\mathrm c$, required for shear, only locally within a shear band, while the overall background porosity $\approx \phi_\mathrm{qs}<\phi_\mathrm{c}$ is below the critical threshold. Upon increasing grain inertia, the background porosity rises along with Eq.~(\ref{phi-I}) (dashed line) and reaches the critical porosity for $I=I_c$. For $I>I_c$, porosity is uniform and larger than $\phi_\mathrm c$, allowing for distributed shear in the entire volume. Shaded areas and yellow lines schematically represent sheared zones and displacement profiles, respectively. (b) Numerical data for porosity inside (asterisks) and outside (circles) localized-shear zones for RBC and $H=96$. The inferred critical values are $\phi_\mathrm{c}-\phi_\mathrm{qs}=3.2\cdot 10^{-3}$ and $I_\mathrm c=6\cdot 10^{-3}$.}
\label{fig:cartoon}
\end{figure}

The model is schematically depicted in Fig.~\ref{fig:cartoon}a. In the quasistatic regime, porosity is constrained to have a bimodal distribution: $\phi(y)=\phi_\mathrm{c}$ inside a shear band, as required for shear to occur, and a lower value $\phi(y)=\phi_\mathrm{qs}$ outside the shear band. The background porosity, outside the shear band, is identified with the global, depth-averaged porosity, $\phi_\mathrm{qs}$, because the thickness of the shear band is assumed negligible compared to the thickness of the layer. The background porosity attains the lowest permitted value in quasistatic shear, consistent with the requirement to minimize work associated with volume change. On increasing the inertial number, the background porosity increases following the dashed line representing the dilatancy law, Eq.~(\ref{phi-I}). As soon as the background porosity equals the (critical) porosity inside the shear band, shear becomes uniformly distributed over the layer as the whole layer is uniformly dilated at the critical porosity. This moment localization vanishes. The transition occurs at the inertial number 
\begin{equation}
\label{I_c}
I_\mathrm{c}=\phi^{-1}(\phi_\mathrm{c}) \,,
\end{equation} 
where $\phi^{-1}(\phi)$ is the inverse function to Eq.~(\ref{phi-I}).

Figure \ref{fig:cartoon}b shows numerical data used to estimate $\phi_\mathrm{c}$ and $I_\mathrm{c}$. Red asterisks and blue circles represent porosities inside and outside of a shear band, respectively. The shear band porosity is calculated as the average porosity over zones where local shear rate is larger than the average shear rate, \textit{i.e.,} $\dot{\gamma}(y)H/V>1$. Analogously, background porosity is calculated as the average porosity in zones of reduced shear rate, $\dot{\gamma}(y)H/V<1$. Resulting porosities inside and outside of a shear band have consistent values across various setups for $I<10^{-3}$ with averages: $\phi_\mathrm{c} = 0.1886\pm0.0009$ for the shear band porosity (horizontal red line) and $\phi = 0.1854\pm0.0011$ for the background porosity (horizontal blue line). The latter value is identical with $\phi_\mathrm{qs}$ obtained from the fit to global porosity data, Eq.~(\ref{phi-I}), along with the assumption of small shear band thickness. For $I>10^{-3}$, the background porosity rises following Eq.~(\ref{phi-I}) (dashed line) and eventually exceeds $\phi_\mathrm{c}$. The inertial number at the transition of localized-distributed shear was calculated from Eq.~(\ref{I_c}) using the average difference between shear-band and background porosities under $I<10^{-3}$ (\textit{i.e.,} the difference between red and blue lines) 
\begin{equation}
\label{phi_c}
\phi_\mathrm{c} - \phi_\mathrm{qs} = (3.2\pm0.7)\cdot 10^{-3}  \to  I_\mathrm{c} =(6\pm1)\cdot 10^{-3}\,.
\end{equation} 



The inertial number $I_\mathrm{c} =6\cdot 10^{-3}$, corresponding to the transition between the localized and the distributed shear regimes as derived from the simple model, is close to the transition $I=1\cdot 10^{-3}$ inferred from the results of numerical simulations (Fig.~\ref{fig:RBC}). It should be noted that the observed transition is not sharp. Instead, the contrast between high and low shear-rate zones fades gradually in the range $10^{-4}<I<10^{-2}$ as the background porosity approaches $\phi_\mathrm{c}$. The barrier $\phi_\mathrm{c}-\phi(I)$ for shear to commence in the background diminishes with growing $I$ and may be overcome by fluctuations of $\phi$. This leads to formation of widespread zones of enhanced shear that are intermittent in time. The threshold $I_\mathrm{c} =6\cdot 10^{-3}$ should be considered as the condition leading to uniformly distributed shear with little fluctuations in space and time.

To sum up, the transition between the localized and the distributed shear regimes coincides with the transition between the quasistatic and the inertial regimes. The latter is identified with $I=10^{-3}$ when the friction coefficient (Fig.~\ref{fig:RBC}e) and global porosity (Fig.~\ref{fig:phi}) depart from their constant, rate-independent values. This is a simple consequence of the fact that porosity, which governs the localized-distributed shear transition, is controlled by inertial number. In the quasistatic regime of low $I$, the shearing layer is in the most compacted and ordered state allowing shear, in which $ \phi$ is minimum. The layer dilates only locally within a shear band to minimize the global volumetric strain $\mathrm d \epsilon_V$, \textit{cf.} Eq.~(\ref{w}). In addition, $\tau_f$ (and thus $\mu_f$) is minimized by grains rolling on a single plane \cite{Makedonska2011} and is rate-independent due to the independency of contact friction forces of velocity, $F_{t_{ij}} = \mu_{g}  F_{n_{ij}}$, characterizing enduring contacts. On the other hand, once shear rate is increased upon the transition to the inertial regime, grain inertia becomes important and both $\tau_f$ and $\epsilon_V$ grow. Inertial effects lead to stronger collisions and agitation. This in turn generates larger interparticle penetrations and therefore larger dissipation $\tau_f \mathrm d \gamma$, while agitation leads to more uniform and enhanced porosity. As a result, $\mu$ and $\phi$ increase hand in hand from the onset of the inertial regime.

A similar qualitative description was provided in \cite{Marone1990,Lyu2019}. The relationship between friction and porosity can be rationalized based on changes in shear band width. The shear band width increases with slip rate (and thereby with $I$), as grain inertia and fracture become increasingly active and mobilize particles several grain diameters from the slip surface. This leads to slip rate dependent porosity and also friction, because dilatancy associated with a wider shear band requires additional work, according to Eq. (\ref{w}). 

Since the transition into the distributed shear regime coincides with frictional strengthening due to inertial effects, such transition may be completely missing if localization is accompanied by frictional weakening that overcompensates the inertial strengthening. Effects such as shear heating or grain comminution inside a shear band may produce sufficient weakening and persistent localization \cite{rice2006,DiToro2014,MAIR2008,Sulem1990,Morgan1999a}. 
The present micro-mechanical model (Sect.~\ref{sec:numerics}) does not include additional weakening effects. One of the consequences is that a shear band is not locked at its incipient position but migrates through the layer.

Previous works also considered the connection between strain localization and rheology. A possible rationalization for the onset of shear banding is non-monotonic rheology, as observed by \textit{e.g.,} \cite{DeGiuli2017,Dijksman2011,Kuwano2013}. In particular, DeGiuli and Wyart considered a similar simulated system of disks under RBC and observed a non-monotonic $\mu(I)$ rheology (see their Fig.~1C) with the transition to a positively sloped (and hence strengthening) friction for an inertial number around $10^{-3}$. Barker and Gray \cite{barker_gray_2017} performed stability analysis of the classical form of $\mu(I)$ relationship \cite{Jop} and found that small wavelength perturbations grow unstably for $I<4\cdot 10^{-3}$, using parameters for spherical glass beads. Again, this critical inertial number is very similar to that observed here. Surprisingly, the transition to unstable regime occurs also for incompressible flow and monotonic friction law, so neither dilatancy nor frictional weakening is required to rationalize onset of instabilities at the continuum scale. Barker and Gray proposed a modified friction law that provided regularization of the incompressible granular flow for inertial numbers down to $0$, and its results matched several experimental observations. On the other hand, the regularized friction law lacked shear localization within shear bands at low $I$. 

\added{One of the most successful approaches in constitutive modeling of granular media are non-local models \cite{Koval2009,Bocquet2009,Kamrin2012,Kamrin2015,Pouliquen2009}, which postulate a distance-dependent drop in yield stress induced by diffusion of a dynamic variable (\textit{e.g.,} fluidity) from slip zones. These models have successfully predicted flow profiles in geometries with inhomogeneous stress distributions. However, for the simple shear geometry considered here, stress is uniform and so is the predicted shear rate. Presence of a shear band would perhaps require a perturbation of the flow law that would render the uniform solution for shear rate unstable. Note that porosity within the non-local fluidity model \cite{Kamrin2012,Kamrin2015} is dictated by the inertial number, similarly to Eq.~\ref{phi-I}. As a result, porosity is a constant independent of stresses in the quasistatic regime, in contrast to the classical localization models \cite{Rice1975}, which require dilatant hardening or weakening. Another possibility for including shear banding within non-local models would correspond to a first-order phase transition scenario, \textit{i.e.}, the spatial coexistence between two states of different fluidity for the same shear stress, as suggested in \cite{Bocquet2009}.}

Future work is required to reveal the connection between instabilities formed in the continuum models and shear bands.

\section{Summary}
In this paper we presented results from a series of 2D steady-state simulations of shearing granular layers between rough walls, under different inertial numbers $I$, normal stresses $N$, and layer thicknesses $H$. Two different boundary conditions were tested: constant shear-strain rate (RBC) and constant applied shear stress (SBC). The simulation results indicate that shear states can be categorized into two main regimes:
\begin{enumerate}
\item At low $I < 10^{-3}$ (slow shear rate, large normal stress or small grain size) quasistatic deformation prevails, with porosity and friction attaining minimum values. This regime occurs only under RBC. Quasistatic shear exhibits intermittent localized shear states, \textit{i.e.,} shear concentrated within shear bands that migrate randomly with time throughout the depth of the layer. This migration is associated with lack of any additional frictional weakening mechanisms in the present model; weakening effects such as shear heating or grain breaking may lock localization onto a persistent position. 
\item For high $I > 10^{-3}$ both friction and mean porosity increase monotonically with $I$ as a result of stronger dissipation and agitation due to growing grain inertia. This inertial regime exhibits distributed flow with no shear bands. The inertial regime occurs for both RBC and SBC. In the latter case, friction depends not only on $I$, but also on \replaced{the contact stiffness number}{grain stiffness} and layer thickness.
\end{enumerate}

The transition between the localized and the distributed shear regimes is explained via the physics of dilatancy and its connection to energy \cite{Marone1990,Rowe1962,Frank1965}: In order to shear, compacted granular media must dilate and attain a porosity larger than or equal to $\phi_\mathrm{c}$, the critical state value \cite{Reynolds1885,Rowe1962,Lambe1969,Desrues1996,Spiers2007}. According to Eq.~(\ref{phi-I}) and Fig.~\ref{fig:phi}, porosity is an increasing function of $I$. Thus, at high enough $I$ porosity exceeds the critical state value and shear can occur within the entire volume of the flow. Dilation needs to only occur in the quasistatic shear, where the overall porosity is below the critical porosity. In this case, dilation within a localized thin shear band minimizes work associated against the applied normal stress, \textit{cf.} Eq.~(\ref{w}), and also minimizes friction \cite{Rowe1962,Marone1990}. This is the reason why localization only emerges in quasistatic shear, and why the transition to distributed shear coincides with the transition to the inertial regime of enhanced friction and porosity.

Despite the common thinking that constant rate and constant stress boundary conditions are equivalent, the quasistatic shear regime appears only under constant rate boundary conditions, while constant applied shear stress always leads to the inertial regime. This can be rationalized using the observation of enhanced shear-stress fluctuations under RBC, Fig.~\ref{fig:fluct}. In the quasistatic regime, these fluctuations are associated with buildup and collapse of force chains \cite{Muhlhaus1987}. Shear resistance is momentarily high when there is configurational ``jamming'' (\textit{e.g.}, buildup of a force chain) and low when the system unjams and accelerates (\textit{e.g.}, collapse of the force chain). On average this gives the low value of quasistatic friction during slow RBC shear. If the shear-stress fluctuations are reduced, as in SBC, we must choose a higher value of mean shear stress, sufficient to unjam the system, or else the system will stop the moment it encounters a strong enough configuration. Therefore, friction required to sustain deformation under SBC slightly exceeds the quasistatic value.

Friction coefficient is a function of the inertial number for RBC. For SBC, friction is also dominantly controlled by the inertial number, but some deviation is observed relative to the $\mu(I)$ dependence measured for RBC. This deviation increases with \replaced{normal stress}{softness (inverse stiffness) of grains}. \replaced{An increase in normal stress leads to an increase in compressive strain of grains. Squashing of grains during configurational jamming accommodates a part of the layer's recoverable elastic energy.}{The softness can be expressed through the non-dimensional ratio of applied normal stress and grain's elastic modulus, $N/E$, which is proportional to the average compressive strain of grains, assuming the linear elasticity model governing contacts between grains. Soft grains are squashed during configurational jamming; they accommodate a part of the layer's strain by recoverable elastic compression while reducing layer's dilatancy compared to rigid grains.} After the jammed state, the elastic energy is released and converted into grains' inertia. As a result, \replaced{larger normal stresses result in}{softer grains (larger $N/E$) display} larger $I$ at a given friction coefficient (Fig.~\ref{fig:SBC}e) and smaller fluctuations of shear stress at a given $I$ (Fig.~\ref{fig:fluct}). For RBC, jammed configurations are associated with generation of larger shear stress compared to SBC. This leads to increased dissipation and friction relative to SBC for the same $I$. Consequently, increased wear and grain comminution is expected under RBC.


We finally discuss the consequences of this work for natural granular flows. Most landslides and faults move in the quasistatic (low $I$) regime, even if intuitively we think of them as inertial since they are rapid. For example, the inertial number of a 100\,m deep landslide that moves at the rapid velocity of 1\,ms$^{-1}$, and has grains of diameter of 1\,cm, is $I \simeq 10^{-6}$. A fault that is buried 1\,km deep, has a gouge zone 10\,cm thick, grains of $10^{-4}$\,m diameter and moves at the seismic speed of 1\,ms$^{-1}$, has $I\simeq10^{-5}$. In contrast, relatively thin debris flows, with thicknesses up to a few tens of meters and with large boulders of 0.1-1\,m diameter, are in the inertial non-localized regimes. Such debris flows indeed tend to bounce around, \replaced{\textit{e.g.,} the famous Elm landslide was described to move like a ``herd of galloping sheep"}{and flow like a stream of shooting debris down a hill slope. This scenario occurred for example in the Elm landslide} \cite{Hsu1978}. Also the recent work of \cite{Li2021} showed that shallow slides tend to flow with a distributed deformation, while deep slides localize deformation. The present work \replaced{implies}{implicates} that many slides and faults, especially deep ones, will move quasistatically and as a consequence will localize shear. Localization will occur first via the mechanism of minimization of dilation, as described above, and after a little strain this localization will lock its position via a variety of mechanisms including grain breakage, shear heating, pore pressure effects and mineral alterations, \textit{e.g.,} \cite{Sulem2013,Sulem2012,rice2006,GOREN2009,Einav2015,BenZeev2020,BenZeev2017}. Thus localization is the rule, rather than the exception in deep geophysical motion, while distributed shear is restricted to motion within the \replaced{upper}{top} 10 -- 20 \deleted{upper }meters near the surface.

\begin{acknowledgements}
S.P., T.T. and M.S. are grateful for the support of Grant No. 19-21114Y from the Czech Science Foundation (GA CR). E.A. acknowledges the support of ISF Grant No. 910/17. Computational resources were supplied by the project ``e-Infrastruktura CZ" (e-INFRA LM2018140) provided within the program Projects of Large Research, Development and Innovations Infrastructures.
\end{acknowledgements}

\section*{Author contributions}
S.P. designed the test of boundary conditions, contributed to the data analysis, interpreted the results and wrote the article. T.T. and M.S. conducted the numerical simulations and contributed to the data analysis. E.A. designed the theoretical model for porosity control over strain localization and contributed to writing of the article.

\bibliographystyle{spphys}       
\bibliography{my_present} 

\begin{thebibliography}{10}
\providecommand{\url}[1]{{#1}}
\providecommand{\urlprefix}{URL }
\expandafter\ifx\csname urlstyle\endcsname\relax
  \providecommand{\doi}[1]{DOI \discretionary{}{}{}#1}\else
  \providecommand{\doi}{DOI \discretionary{}{}{}\begingroup
  \urlstyle{rm}\Url}\fi

\bibitem{Scholz1987}
C.H. Scholz, Geology \textbf{15}(6), 493 (1987).
\newblock \doi{$10.1130/0091-7613(1987)15<493:WAGFIB>2.0.CO;2$}

\bibitem{Chester1993}
F.M. Chester, J.P. Evans, R.L. Biegel, Journal of Geophysical Research: Solid
  Earth \textbf{98}(B1), 771 (1993).
\newblock \doi{10.1029/92JB01866}

\bibitem{Billi2005}
A.~Billi, J. Struct. Geol. \textbf{27}, 1823 (2005).
\newblock \doi{10.1016/j.jsg.2005.05.013}

\bibitem{Arboleya1995}
M.L. Arboleya, T.~Engelder, J. Struct. Geol. \textbf{17}, 519 (1995).
\newblock \doi{10.1016/0191-8141(94)00079-F}

\bibitem{Cashman2000}
S.~Cashman, K.~Cashman, Geology \textbf{28}, 111 (2000).
\newblock \doi{$10.1130/0091-7613(2000)28<111:CADFIU>2.0.CO;2$}

\bibitem{Hayman2004}
N.W. Hayman, B.A. Housen, T.T. Cladouhos, K.~Livi, Journal of Geophysical
  Research: Solid Earth \textbf{109}(B5) (2004).
\newblock \doi{10.1029/2003JB002902}

\bibitem{Boullier2009}
A.M. Boullier, E.C. Yeh, S.~Boutareaud, S.R. Song, C.H. Tsai, Geochemistry,
  Geophysics, Geosystems \textbf{10}(3) (2009).
\newblock \doi{10.1029/2008GC002252}

\bibitem{Shalev2013}
S.~Siman-Tov, E.~Aharonov, A.~Sagy, S.~Emmanuel, Geology \textbf{41}(6), 703
  (2013).
\newblock \doi{10.1130/G34087.1}

\bibitem{Smeraglia2017}
L.~Smeraglia, A.~Billi, E.~Carminati, A.~Cavallo, G.~Di~Toro, E.~Spagnuolo,
  F.~Zorzi, Scientific Reports \textbf{7}(1), 664 (2017).
\newblock \doi{10.1038/s41598-017-00717-4}

\bibitem{Logan1979}
J.M. Logan, M.~Friedman, N.~Higgs, C.~Dengo, T.~Shimamoto, U.S. Geol. Surv.
  Open File Rep. \textbf{79--1239}, 305 (1979)

\bibitem{Marone1990}
C.~Marone, C.B. Raleigh, C.H. Scholz, J. Geophys. Res. \textbf{95}, 7007
  (1990).
\newblock \doi{10.1029/JB095iB05p07007}

\bibitem{LOGAN1992}
J.~Logan, C.~Dengo, N.~Higgs, Z.~Wang, in \emph{Fault Mechanics and Transport
  Properties of Rocks}, \emph{International Geophysics}, vol.~51, ed. by
  B.~Evans, T.~fong Wong (Academic Press, 1992), pp. 33 -- 67.
\newblock \doi{https://doi.org/10.1016/S0074-6142(08)62814-4}

\bibitem{Beeler1996}
N.M. Beeler, T.E. Tullis, M.L. Blanpied, J.D. Weeks, Journal of Geophysical
  Research: Solid Earth \textbf{101}(B4), 8697 (1996).
\newblock \doi{10.1029/96JB00411}

\bibitem{Spiers2007}
A.R. Niemeijer, C.J. Spiers, J. Geophys. Res. \textbf{112}, S78 (2007).
\newblock \doi{http://dx.doi.org/10.1016/j.ijggc.2012.09.018}

\bibitem{Reches2010}
Z.~Reches, D.~Lockner, Nature \textbf{467}, 452 (2010).
\newblock \doi{10.1038/nature09348}

\bibitem{DiToro2014}
B.P. Proctor, T.M. Mitchell, G.~Hirth, D.~Goldsby, F.~Zorzi, J.D. Platt,
  G.~Di~Toro, Journal of Geophysical Research: Solid Earth \textbf{119}(11),
  8107 (2014).
\newblock \doi{10.1002/2014JB011057}

\bibitem{Mitchell2016}
E.K. Mitchell, Y.~Fialko, K.M. Brown, J. Geophys. Res. Solid Earth
  \textbf{121}, 6932 (2016)

\bibitem{Mora1999}
P.~Mora, D.~Place, Geophysical Research Letters \textbf{26}(1), 123 (1999).
\newblock \doi{10.1029/1998GL900231}

\bibitem{Aharonov2002}
E.~Aharonov, D.~Sparks, Phys. Rev. E \textbf{65}, 051302 (2002)

\bibitem{Morgan1999a}
J.K. Morgan, M.S. Boettcher, Journal of Geophysical Research: Solid Earth
  \textbf{104}(B2), 2703 (1999).
\newblock \doi{10.1029/1998JB900056}

\bibitem{MAIR2008}
K.~Mair, S.~Abe, Earth and Planetary Science Letters \textbf{274}(1), 72
  (2008).
\newblock \doi{https://doi.org/10.1016/j.epsl.2008.07.010}

\bibitem{Li2021}
K.~Li, Y.F. Wang, Q.W. Lin, Q.G. Cheng, Y.~Wu, Landslides \textbf{18}, 1779
  (2021).
\newblock \doi{10.1007/s10346-020-01607-z}

\bibitem{Ben-Zion2003}
Y.~Ben-Zion, C.G. Sammis, pure and applied geophysics \textbf{160}(3), 677
  (2003)

\bibitem{Marone1991}
C.~Marone, PAGEOPH \textbf{137}, 409 (1991).
\newblock \doi{10.1007/BF00879042}

\bibitem{Brace1972}
W.F. Brace, Tectonophysics \textbf{14}, 189 (1972)

\bibitem{Ikari2011}
M.~Ikari, C.~Marone, D.~Saffer, Geology \textbf{39}, 83 (2011).
\newblock \doi{10.1130/G31416.1}

\bibitem{Shimamoto1986}
T.~Shimamoto, Science \textbf{231}, 711 (1986)

\bibitem{French2016}
M.~French, W.~Zhu, J.~Banker, Geophys. Res. Lett. \textbf{43}, 4330 (2016).
\newblock \doi{10.1002/2016GL068893}

\bibitem{Wu2013}
W.~Wu, Y.~Zou, X.~Li, J.~Zhao, Review of Scientific Instruments \textbf{85},
  093902 (2014).
\newblock \doi{10.1063/1.4894207}

\bibitem{Marone2009}
J.~Samuelson, D.~Elsworth, C.~Marone, J. Geophys. Res. \textbf{114}, B12404
  (2009).
\newblock \doi{10.1029/2008JB006273}

\bibitem{Faulkner2018}
D.R. Faulkner, C.~Sanchez-Roa, C.~Boulton, S.A.M. den Hartog, Journal of
  Geophysical Research: Solid Earth \textbf{123}, 1 (2018).
\newblock \doi{10.1002/2017JB015130}

\bibitem{GOREN2009}
L.~Goren, E.~Aharonov, Earth and Planetary Science Letters \textbf{277}(3), 365
   (2009).
\newblock \doi{https://doi.org/10.1016/j.epsl.2008.11.002}

\bibitem{Chester1994}
F.M. Chester, J. Geophys. Res. Solid Earth \textbf{99}, 7247 (1994)

\bibitem{Aharonov2018}
E.~Aharonov, C.H. Scholz, Journal of Geophysical Research: Solid Earth
  \textbf{123}(2), 1591 (2018).
\newblock \doi{10.1002/2016JB013829}

\bibitem{Frye2002}
K.M. Frye, C.~Marone, Journal of Geophysical Research: Solid Earth
  \textbf{107}(B11), ETG 11 (2002).
\newblock \doi{https://doi.org/10.1029/2001JB000654}.
\newblock
  \urlprefix\url{https://agupubs.onlinelibrary.wiley.com/doi/abs/10.1029/2001JB000654}

\bibitem{Rice1975}
J.W. Rudnicki, J.R. Rice, J. Mech. Phys. Solids \textbf{23}, 371 (1975)

\bibitem{Rice1976}
J.R. Rice, in \emph{Theoretical and Applied Mechanics (Proceedings of the 14th
  International Congress on Theoretical and Applied Mechanics)}, ed. by W.T.
  Koiter (North-Holland Publishing, 1976), pp. 207--220

\bibitem{VARDOULAKIS1976}
I.~Vardoulakis, Mechanics Research Communications \textbf{3}(3), 209  (1976).
\newblock \doi{https://doi.org/10.1016/0093-6413(76)90014-8}

\bibitem{Vardoulakis1980}
I.~Vardoulakis, International Journal for Numerical and Analytical Methods in
  Geomechanics \textbf{4}(2), 103 (1980).
\newblock \doi{10.1002/nag.1610040202}

\bibitem{Muhlhaus1987}
H.B. Muhlhaus, I.~Vardoulakis, G\'eotechnique \textbf{37}(3), 271 (1987).
\newblock \doi{10.1680/geot.1987.37.3.271}

\bibitem{Sulem1990}
J.~Sulem, I.~Vardoulakis, Acta Mechanica \textbf{83}, 195 (1990).
\newblock \doi{10.1007/BF01172981}

\bibitem{LARSSON1996}
R.~LARSSON, K.~RUNESSON, K.~AXELSSON, International Journal for Numerical and
  Analytical Methods in Geomechanics \textbf{20}(11), 771 (1996).
\newblock
  \doi{$10.1002/(SICI)1096-9853(199611)20:11<771::AID-NAG847>3.0.CO;2-M$}

\bibitem{Weir2003}
G.~Weir, R.~Young, International Journal for Numerical and Analytical Methods
  in Geomechanics \textbf{27}(15), 1299 (2003).
\newblock \doi{10.1002/nag.322}

\bibitem{Einav2006}
I.~Einav, M.~Randolph, G\'eotechnique \textbf{56}(7), 501 (2006).
\newblock \doi{10.1680/geot.2006.56.7.501}

\bibitem{Vardoulakis1995}
I.~Vardoulakis, J.~Sulem, \emph{Bifurcation Analysis in Geomechanics} (Taylor
  \& Francis, Oxon, 1995)

\bibitem{rice2006}
J.R. Rice, J. Geophys. Res. \textbf{111}, B05311 (2006).
\newblock \doi{10.1029/2005JB004006}

\bibitem{Sulem2009}
J.~Sulem, V.~Famin, Journal of Geophysical Research: Solid Earth
  \textbf{114}(B3), B03309 (2009).
\newblock \doi{10.1029/2008JB006004}

\bibitem{Veveakis2011}
J.~Sulem, I.~Stefanou, E.~Veveakis, Granular Matter \textbf{13}, 261 (2011).
\newblock \doi{10.1007/s10035-010-0244-1}

\bibitem{Sulem2012}
N.~Brantut, J.~Sulem, Journal of Applied Mechanics \textbf{79}(3) (2012).
\newblock \doi{10.1115/1.4005880}

\bibitem{Sulem2013}
M.~Veveakis, I.~Stefanou, S.~J., Geotechnique Letters \textbf{3}, 31 (2013).
\newblock \doi{10.1680/geolett.12.00063}

\bibitem{Rice2014}
J.R. Rice, J.W. Rudnicki, J.D. Platt, Journal of Geophysical Research: Solid
  Earth \textbf{119}(5), 4311 (2014).
\newblock \doi{10.1002/2013JB010710}

\bibitem{Desrues1996}
J.~Desrues, R.~Chambon, M.~Mokni, F.~Mazerolle, G\'eotechnique \textbf{46}(3),
  529 (1996).
\newblock \doi{10.1680/geot.1996.46.3.529}

\bibitem{Lambe1969}
T.W. Lambe, R.V. Whitman, \emph{Soil Mechanics} (John Wiley, New York, 1969)

\bibitem{GDRMiDi}
G.D.R. MiDi, Eur. Phys. J. E \textbf{14}, 341 (2004)

\bibitem{Forterre2008}
Y.~Forterre, O.~Pouliquen, Annu. Rev. Fluid Mech. \textbf{40}, 1 (2008)

\bibitem{daCruz2005}
F.~da~Cruz, S.~Emam, M.~Prochnow, J.N. Roux, F.~Chevoir, Phys. Rev. E
  \textbf{72}, 021309 (2005)

\bibitem{Singh2015}
A.~Singh, V.~Magnanimo, K.~Saitoh, S.~Luding, New J. Phys. \textbf{17}, 043028
  (2015).
\newblock \doi{10.1088/1367-2630/17/4/043028}

\bibitem{Parez2015}
S.~Parez, E.~Aharonov, Front. Phys. \textbf{3}, 80 (2015).
\newblock \doi{10.3389/fphy.2015.00080}

\bibitem{Parez2016}
S.~Parez, E.~Aharonov, R.~Toussaint, Phys. Rev. E \textbf{93}, 042902 (2016).
\newblock \doi{10.1103/PhysRevE.93.042902}

\bibitem{Shojaaee2012a}
Z.~Shojaaee, J.N. Roux, F.~Chevoir, D.E. Wolf, Phys. Rev. E \textbf{86}, 011301
  (2012)

\bibitem{DeGiuli2017}
E.~DeGiuli, M.~Wyart, Proceedings of the National Academy of Sciences
  \textbf{114}(35), 9284 (2017).
\newblock \doi{10.1073/pnas.1706105114}.
\newblock \urlprefix\url{https://www.pnas.org/content/114/35/9284}

\bibitem{Dijksman2011}
J.A. Dijksman, G.H. Wortel, L.T.H. van Dellen, O.~Dauchot, M.~van Hecke, Phys.
  Rev. Lett. \textbf{107}, 108303 (2011).
\newblock \doi{10.1103/PhysRevLett.107.108303}.
\newblock
  \urlprefix\url{https://link.aps.org/doi/10.1103/PhysRevLett.107.108303}

\bibitem{Kuwano2013}
O.~Kuwano, R.~Ando, T.~Hatano, Geophysical Research Letters \textbf{40}(7),
  1295 (2013).
\newblock \doi{https://doi.org/10.1002/grl.50311}.
\newblock
  \urlprefix\url{https://agupubs.onlinelibrary.wiley.com/doi/abs/10.1002/grl.50311}

\bibitem{Pouliquen_book}
B.~Andreotti, Y.~Forterre, O.~Pouliquen, \emph{Granular Media: Between Fluid
  and Solid} (Cambridge University Press, Cambridge, England, 2013)

\bibitem{Melosh1979}
H.J. Melosh, Journal of Geophysical Research: Solid Earth \textbf{84}(B13),
  7513 (1979).
\newblock \doi{https://doi.org/10.1029/JB084iB13p07513}.
\newblock
  \urlprefix\url{https://agupubs.onlinelibrary.wiley.com/doi/abs/10.1029/JB084iB13p07513}

\bibitem{barker_gray_2017}
T.~Barker, J.M.N.T. Gray, Journal of Fluid Mechanics \textbf{828}, 5–32
  (2017).
\newblock \doi{10.1017/jfm.2017.428}

\bibitem{Clark2018}
A.H. Clark, J.D. Thompson, M.D. Shattuck, N.T. Ouellette, C.S. O'Hern, Phys.
  Rev. E \textbf{97}, 062901 (2018).
\newblock \doi{10.1103/PhysRevE.97.062901}.
\newblock \urlprefix\url{https://link.aps.org/doi/10.1103/PhysRevE.97.062901}

\bibitem{Frank1965}
F.C. Frank, Reviews of Geophysics \textbf{3}(4), 485 (1965).
\newblock \doi{10.1029/RG003i004p00485}

\bibitem{Reynolds1885}
O.~Reynolds, Phylos. Mag. Ser. 5 \textbf{20}, 469 (1885)

\bibitem{DEM}
P.A. Cundall, O.D. Strack, G\'eotechnique \textbf{29}, 47 (1979)

\bibitem{Frenkel2002}
D.~Frenkel, B.~Smit, \emph{Understanding molecular simulations} (Academic
  Press, San Diego, 2002)

\bibitem{Shojaaee2012b}
Z.~Shojaaee, L.~Brendel, J.~Torok, D.E. Wolf, Phys. Rev. E \textbf{86}, 011302
  (2012)

\bibitem{num_Silbert}
L.E. Silbert, D.~Ertas, G.S. Grest, T.C. Halsey, D.~Levine, S.J. Plimpton,
  Phys. Rev. E \textbf{64}, 051302 (2001)

\bibitem{deCoulomb2017}
A.~Favier~de Coulomb, M.~Bouzid, P.~Claudin, E.~Cl\'ement, B.~Andreotti, Phys.
  Rev. Fluids \textbf{2}, 102301 (2017).
\newblock \doi{10.1103/PhysRevFluids.2.102301}.
\newblock
  \urlprefix\url{https://link.aps.org/doi/10.1103/PhysRevFluids.2.102301}

\bibitem{Koval2009}
G.~Koval, J.N. Roux, A.~Corfdir, F.~Chevoir, Phys. Rev. E \textbf{79}, 021306
  (2009)

\bibitem{Bocquet2009}
L.~Bocquet, A.~Colin, A.~Ajdari, Phys. Rev. Lett. \textbf{103}, 036001 (2009).
\newblock \doi{10.1103/PhysRevLett.103.036001}.
\newblock
  \urlprefix\url{https://link.aps.org/doi/10.1103/PhysRevLett.103.036001}

\bibitem{Kamrin2012}
K.~Kamrin, G.~Koval, Phys. Rev. Lett. \textbf{108}, 178301 (2012)

\bibitem{Kamrin2015}
K.~Kamrin, D.L. Henann, Soft Matter \textbf{11}, 179 (2015)

\bibitem{Pouliquen2009}
O.~Pouliquen, Y.~Forterre, Phil. Trans. R. Soc. A \textbf{367}, 5091 (2009).
\newblock \doi{10.1098/rsta.2009.0171}

\bibitem{Jop}
P.~Jop, Y.~Forterre, O.~Pouliquen, Nature \textbf{441}, 727 (2006)

\bibitem{Rowe1962}
P.W. Rowe, G.I. Taylor, Proceedings of the Royal Society of London. Series A.
  Mathematical and Physical Sciences \textbf{269}(1339), 500 (1962).
\newblock \doi{10.1098/rspa.1962.0193}

\bibitem{Makedonska2011}
N.~Makedonska, D.W. Sparks, E.~Aharonov, L.~Goren, J. Geophys. Res.
  \textbf{116}, B09302 (2011)

\bibitem{Lyu2019}
Z.~Lyu, J.~Rivière, Q.~Yang, C.~Marone, Tectonophysics \textbf{763}, 86
  (2019).
\newblock \doi{https://doi.org/10.1016/j.tecto.2019.04.010}.
\newblock
  \urlprefix\url{https://www.sciencedirect.com/science/article/pii/S0040195119301325}

\bibitem{Hsu1978}
K.J. Hsu, in \emph{Rockslides and Avalanches}, ed. by B.~Voight (Elsevier,
  Amsterdam, 1978), pp. 71--93

\bibitem{Einav2015}
F.~Guillard, P.~Golshan, L.~Shen, J.R. Valdes, I.~Einav, Nature Physics
  \textbf{11}(10), 835 (2015).
\newblock \doi{10.1038/nphys3424}

\bibitem{BenZeev2020}
S.~Ben-Zeev, E.~Aharonov, R.~Toussaint, S.~Parez, L.~Goren, Phys. Rev. Fluids
  \textbf{5}, 054301 (2020).
\newblock \doi{10.1103/PhysRevFluids.5.054301}.
\newblock
  \urlprefix\url{https://link.aps.org/doi/10.1103/PhysRevFluids.5.054301}

\bibitem{BenZeev2017}
S.~Ben-Zeev, L.~Goren, S.~Parez, R.~Toussaint, C.~Clément, E.~Aharonov,
  \emph{The Combined Effect of Buoyancy and Excess Pore Pressure in
  Facilitating Soil Liquefaction} (2017), pp. 107--116.
\newblock \doi{10.1061/9780784480779.013}

\end{thebibliography}


\end{document}